\global\def\draftcontrol{0}

   \def\versionno{Gluon Scattering  from Sugra}

\catcode`\@=11

\expandafter\ifx\csname draftcontrol\endcsname\relax\global\def\draftcontrol{0}
\fi

{\count255=\time\divide\count255 by 60
\xdef\hourmin{\number\count255}
\multiply\count255 by-60\advance\count255 by\time
\xdef\hourmin{\hourmin:\ifnum\count255<10 0\fi\the\count255}}
\def\draftdate{\number\month/\number\day/\number\year\ \ \ \hourmin }

\newcommand\makepapertitle{\par
  \begingroup
    \renewcommand\thefootnote{\@fnsymbol\c@footnote}%
    \def\@makefnmark{\rlap{\@textsuperscript{\normalfont\@thefnmark}}}%
    \long\def\@makefntext##1{\parindent 1em\noindent
            \hb@xt@1.8em{%
                \hss\@textsuperscript{\normalfont\@thefnmark}}##1}%
     \newpage
     \global\@topnum\z@   
     \@makepapertitle
     \thispagestyle{empty}\@thanks
  \endgroup
  \setcounter{footnote}{0}%
  \global\let\thanks\relax
  \global\let\makepapertitle\relax
  \global\let\@makepapertitle\relax
  \global\let\@thanks\@empty
  \global\let\@author\@empty
  \global\let\@date\@empty
  \global\let\@title\@empty
  \global\let\title\relax
  \global\let\author\relax
  \global\let\date\relax
  \global\let\and\relax
  \def\version{\let\version\@version\@gobble}
}
\def\@makepapertitle{%
  \newpage
   \ifnum\draftcontrol=1 {}
   \version\versionno
   \vskip 3em%
   \else
   \hfill\hbox to 3cm {\parbox{4cm}{\@pubnum}\hss}%
   \vskip 3em%
   \fi
   \begin{center}%
   \let \footnote \thanks
     {\LARGE {\@title}}%
     \vskip 1.5em%
     {\normalsize
       \lineskip .5em%
       \begin{tabular}[t]{c}%
         \@author
       \end{tabular}\par}%
     \vskip 1.5em%
     {\@bstract}%
     \end{center}%
     \vskip 1.5em
     \@date%
   \par
}

\gdef\@pubnum{}
\def\pubnum#1{%
  \gdef\@pubnum{#1}}

\gdef\@bstract{}
\def\Abstract#1{%
  \gdef\@bstract{%
   \parbox{\textwidth-0pc}{%
   \centerline{\bf Abstract}\penalty1000%
\noindent
\renewcommand\baselinestretch{1.0}%
{#1}}}
}

\def\ps@paper{\let\@mkboth\@gobbletwo%
     \ifnum\draftcontrol=1
        \def\@oddfoot{\hbox to \textwidth{\tiny \versionno \hfil\tiny\draftdate}%
        \hskip -\textwidth \hbox to \textwidth{\hfil\rm\thepage\hfil}}%
     \else\def\@oddfoot{\hbox to \textwidth{\hfil\rm\thepage\hfil}}
     \fi
     \let\@evenfoot\@oddfoot
}

\def\@version#1{\ifnum\draftcontrol=1
\typeout{}\typeout{#1}\typeout{}
\vskip3mm\centerline{\hbox{\fbox{\normalsize{\tt DRAFT -- #1 -- }
                   {\draftdate}}}}\vskip3mm
\fi}
\let\version\@version
\long\def\eqlabel#1{\ifnum\draftcontrol=1
                    \tag@false  
                    \tag*{(\theequation) \hbox to -0.2cm{\hspace{0cm}\small{#1}\hss}}
                    \refstepcounter{equation}
                    \edef\@currentlabel{\theequation}
                    \ltx@label{#1}          
                    \else
                    \label{#1}
                    \fi
                    }
\let\st@bibitem\@bibitem
\let\st@lbibitem\@lbibitem
\ifnum\draftcontrol=1
  \def\@bibitem#1{%
    \st@bibitem{#1}\a@@label{#1}\ignorespaces}
  \def\@lbibitem[#1]#2{%
    \st@lbibitem[#1]{#2}\a@@label{#2}\ignorespaces}
  \def\a@@label#1{%
    \gdef\a@lab{\smash{\normalfont\small#1}}
    \ifvmode
      \if@inlabel
        \global\setbox\@labels\hbox{%
          \llap{\a@lab\let\a@lab\relax
                \kern\@totalleftmargin\kern\marginparsep}%
          \box\@labels}%
      \fi
    \fi}
\fi

\documentclass[12pt,letterpaper]{article}

\usepackage{amsmath,amssymb,array,calc,rotating,epsfig}
\ifnum\draftcontrol=1
\fi

\renewcommand\baselinestretch{1.25}
\renewcommand\section{\@startsection {section}{1}{\z@}%
                                   {-3.5ex \@plus -1ex \@minus -.2ex}%
                                   {2.3ex \@plus.2ex}%
                                   {\normalfont\large\bfseries}}
\renewcommand\subsection{\@startsection{subsection}{2}{\z@}%
                                   {-3.25ex\@plus -1ex \@minus -.2ex}%
                                   {1.5ex \@plus .2ex}%
                                   {\normalfont\normalsize\bfseries}}
\renewcommand\subsubsection{\@startsection{subsubsection}{3}{\z@}%
                                   {-3.25ex\@plus -1ex \@minus -.2ex}%
                                   {1.5ex \@plus .2ex}%
                                   {\normalfont\normalsize\it}}
\renewcommand\paragraph{\@startsection{paragraph}{4}{\z@}%
                                   {-3.25ex\@plus -1ex \@minus -.2ex}%
                                   {1.5ex \@plus .2ex}%
                                   {\normalfont\normalsize\bf}}

\def\revise#1       {\raisebox{-0em}{\rule{3pt}{1em}}%
                     \marginpar{\raisebox{.5em}{\vrule width3pt\
                     \vrule width0pt height 0pt depth0.5em
                     \hbox to 0cm{\hspace{0cm}{%
                     \parbox[t]{4em}{\raggedright\footnotesize{#1}}}\hss}}}}

\def\del          {\partial}

\def\tr           {\mathop{\rm Tr}}
\def\Re           {{\rm Re\hskip0.1em}}
\def\Im           {{\rm Im\hskip0.1em}}

\def\half{{\frac12}}

\def\sqr#1#2{{\vcenter{\vbox{\hrule height.#2pt
 \hbox{\vrule width.#2pt height#1pt \kern#1pt
 \vrule width.#2pt}\hrule height.#2pt}}}}
\def\square{%
  \mathop{\mathchoice{\sqr{12}{15}}{\sqr{9}{12}}{\sqr{6.3}{9}}{\sqr{4.5}{9}}}}

\newcommand{\fft}[2]{{\frac{#1}{#2}}}

\def\a{\alpha}
\def\b{\beta}
\def\r{\rho}

\def\m{\mu}
\def\g{\gamma}
\def\l{\lambda}
\def\n{\nu}
\def\bn{\bar{\nu}}
\def\bm{\bar{\mu}}

\catcode`\@=12

\usepackage{color}

\begin{document}




\newcommand{\be}{\begin{equation}}
\newcommand{\ee}{\end{equation}}
\newcommand{\beq}{\begin{equation}}
\newcommand{\eeq}{\end{equation}}
\newcommand{\ba}{\begin{eqnarray}}
\newcommand{\ea}{\end{eqnarray}}
\newcommand{\nn}{\nonumber}

\def\vol{\bf vol}
\def\Vol{\bf Vol}
\def\del{{\partial}}
\def\vev#1{\left\langle #1 \right\rangle}
\def\cn{{\cal N}}
\def\co{{\cal O}}
\def\IC{{\mathbb C}}
\def\IR{{\mathbb R}}
\def\IZ{{\mathbb Z}}
\def\RP{{\bf RP}}
\def\CP{{\bf CP}}
\def\Poincare{{Poincar\'e }}
\def\tr{{\rm tr}}
\def\tp{{\tilde \Phi}}
\def\Y{{\bf Y}}
\def\te{\theta}
\def\bX{\bf{X}}

\def\TL{\hfil$\displaystyle{##}$}
\def\TR{$\displaystyle{{}##}$\hfil}
\def\TC{\hfil$\displaystyle{##}$\hfil}
\def\TT{\hbox{##}}
\def\HLINE{\noalign{\vskip1\jot}\hline\noalign{\vskip1\jot}} 
\def\seqalign#1#2{\vcenter{\openup1\jot
  \halign{\strut #1\cr #2 \cr}}}
\def\lbldef#1#2{\expandafter\gdef\csname #1\endcsname {#2}}
\def\eqn#1#2{\lbldef{#1}{(\ref{#1})}%
\begin{equation} #2 \label{#1} \end{equation}}
\def\eqalign#1{\vcenter{\openup1\jot   }}
\def\eno#1{(\ref{#1})}
\def\href#1#2{#2}
\def\half{{1 \over 2}}

\def\ads{{\it AdS}}
\def\adsp{{\it AdS}$_{p+2}$}
\def\cft{{\it CFT}}

\newcommand{\ber}{\begin{eqnarray}}
\newcommand{\eer}{\end{eqnarray}}

\newcommand{\bea}{\begin{eqnarray}}
\newcommand{\eea}{\end{eqnarray}}

\newcommand{\beqar}{\begin{eqnarray}}
\newcommand{\cN}{{\cal N}}
\newcommand{\cO}{{\cal O}}
\newcommand{\cA}{{\cal A}}
\newcommand{\cT}{{\cal T}}
\newcommand{\cF}{{\cal F}}
\newcommand{\cC}{{\cal C}}
\newcommand{\cR}{{\cal R}}
\newcommand{\cW}{{\cal W}}
\newcommand{\eeqar}{\end{eqnarray}}
\newcommand{\lm}{\lambda}\newcommand{\Lm}{\Lambda}
\newcommand{\eps}{\epsilon}


\newcommand{\nonu}{\nonumber}
\newcommand{\oh}{\displaystyle{\frac{1}{2}}}
\newcommand{\dsl}
  {\kern.06em\hbox{\raise.15ex\hbox{$/$}\kern-.56em\hbox{$\partial$}}}
\newcommand{\as}{\not\!\! A}
\newcommand{\ps}{\not\! p}
\newcommand{\ks}{\not\! k}
\newcommand{\D}{{\cal{D}}}
\newcommand{\dv}{d^2x}
\newcommand{\Z}{{\cal Z}}
\newcommand{\N}{{\cal N}}
\newcommand{\Dsl}{\not\!\! D}
\newcommand{\Bsl}{\not\!\! B}
\newcommand{\Psl}{\not\!\! P}
\newcommand{\eeqarr}{\end{eqnarray}}
\newcommand{\ZZ}{{\rm \kern 0.275em Z \kern -0.92em Z}\;}

\def\s{\sigma}
\def\a{\alpha}
\def\b{\beta}
\def\r{\rho}
\def\d{\delta}
\def\g{\gamma}
\def\G{\Gamma}
\def\ep{\epsilon}
\makeatletter \@addtoreset{equation}{section} \makeatother
\renewcommand{\theequation}{\thesection.\arabic{equation}}

\def\be{\begin{equation}}
\def\ee{\end{equation}}
\def\bea{\begin{eqnarray}}
\def\eea{\end{eqnarray}}
\def\m{\mu}
\def\n{\nu}
\def\g{\gamma}
\def\p{\phi}
\def\L{\Lambda}
\def \W{{\cal W}}
\def\bn{\bar{\nu}}
\def\bm{\bar{\mu}}
\def\bw{\bar{w}}
\def\ba{\bar{\alpha}}
\def\bb{\bar{\beta}}


\def\bear{\begin{eqnarray}}
\def\eear{\end{eqnarray}}
\def\nn{\nonumber}

\newcommand\belabel[1]{\begin{equation}\label{#1}}

\newcommand\bearlabel[1]{\begin{eqnarray}\label{#1}}

\newcommand\bra[1]{{\langle {#1}|}}  

\newcommand\ket[1]{{|{#1}\rangle}}  
\newcommand{\braket}[2]{\langle #1|#2\rangle}

\def\defineas{{:=}}

\def\a{\alpha}

\def\b{\beta}

\def\g{\gamma}

\def\r{\rho}

\def\th{\theta}

\def\l{\lambda}
\def\m{\mu}
\def\tL{\tilde L}

\def\cA{\cal A}
\def\cJ{\cal J}
\def\cL{\cal L}
\def\cV{\cal V}
\def\cM{\cal M}
\def\cN{\cal N}
\def\G{\Gamma}
\def\D{\Delta}

\def\N{\mbox{\tiny NS}}

\def\R{\mbox{\tiny R}}

\def\P{\mbox{\tiny P}}

\def\X{\mbox{\tiny X}}
\def\half{\frac{1}{2}}
\def\thalf{\frac{3}{2}}
\def\wt{\widetilde}
\def\nn{\nonumber}

\newcommand{\Tr}{{\rm Tr\,}}
\newcommand{\Der}{{\mathcal D}}
\renewcommand{\tfrac}[2]{{\textstyle{\frac{#1}{#2}}}}
\newcommand{\Det}{\mbox{{\rm Det}}}
\renewcommand{\Re}{\mathrm{Re}\,}
\renewcommand{\Im}{\mathrm{Im}\,}
\newcommand{\pa}{\partial}
\newcommand{\pab}{\bar{\partial}}
\newcommand{\ga}{\gamma}
\newcommand{\gah}{\hat{\gamma}}
\newcommand{\id}{\mathbb I}
\newcommand{\mcj}{\mathcal J}
\newcommand{\gau}[1]{\!\,_{\,_#1}\!A}
\newcommand{\Ws}[1]{W^{#1}}
\newcommand{\gauf}[1]{\!\,_{\,_#1}\!F}
\newcommand{\sig}[1]{\!\,_{\,_#1}\!\Sigma}
\newcommand{\omegs}[1]{\Omega^{#1}}
\newcommand{\pp}[1]{\!\,_{\,_#1}\!\phi}
\newcommand{\oms}[1]{\omega^{#1}}
\newcommand{\ch}[1]{\!\,_{\,_#1}\!\chi}
\newcommand{\chb}[1]{\!\,_{\,_#1}\!\bar{\chi}}
\newcommand{\cht}[1]{\!\,_{\,_#1}\!\tilde{\chi}}
\newcommand{\chtb}[1]{\!\,_{\,_#1}\!\bar{\tilde{\chi}}}
\newcommand{\xis}[1]{\xi^{#1}}
\newcommand{\xibs}[1]{\bar{\xi}^{#1}}
\newcommand{\xits}[1]{\tilde{\xi}^{#1}}
\newcommand{\xitbs}[1]{\bar{\tilde{\xi}}^{#1}}
\renewcommand{\tr}[1]{{\rm tr}_{\,_#1}}
\newcommand{\Th}{\hat{T}}
\newcommand{\Ps}[1]{\!\,_{\,_#1}\!P}

\begin{titlepage}

\version\versionno

\leftline{\tt arXiv:0810.0028}

\vskip -.8cm

\rightline{\small{\tt MCTP-08-63}}
\rightline{\small{\tt }}

\vskip 1.7 cm

\centerline{\bf \Large Coulomb Phase Gluon Scattering}

\vskip .2cm

\centerline{\bf \Large at Strong Coupling}

\vskip 1cm

\centerline{Benjamin A. Burrington$^{1,2}$   and Leopoldo A. Pando Zayas$^{3}$}

\vspace{0.5cm}
\centerline{\it ${}^1$Department of Physics,}
\centerline{\it University of Toronto,}
\centerline{\it Toronto, Ontario, Canada M5S 1A7. }

\vspace{0.5cm}
\centerline{\it ${}^2$School of Physics and Astronomy,}
\centerline{\it The Raymond and Beverly Sackler Faculty of Exact Sciences,} \centerline{\it Tel Aviv University, Ramat Aviv, 69978, Israel. }

\vskip .5cm
\centerline{\it ${}^3$ Michigan Center for Theoretical
Physics}
\centerline{ \it Randall Laboratory of Physics, The University of
Michigan}
\centerline{\it Ann Arbor, MI 48109-1040}

\vspace{1cm}

\begin{abstract}
We calculate corrections to gluon scattering amplitudes in a Coulomb phase using gauge/string duality.  The Coulomb phase considered is a maximal rank breaking of $SU(n_1+n_2)\rightarrow SU(n_1)\times SU(n_2) \times U(1)$.  This problem therefore has 3 scales involved: 1) the scale of the massive fields $M_W$ arising from the spontaneous breaking of the gauge group; 2) The scale of the scattering, characterized by the Mandelstam variables $s,t,u$; 3) The IR regulator $m_{IR}$.  We find corrections in the hard scattering limit $ |s|,|t|,|u|\gg m_{IR}^2 \gg M_W^2$, and also find below threshold corrections with $M_W^2 \gg |s|,|t|,|u|$.  We find that the corrections in the second case are finite, and so are IR regulator independent.
\end{abstract}



\end{titlepage}



\section{Introduction}

The generation of mass for the gauge sector via spontaneous symmetry breaking is a trademark ingredient of the Standard Model. The presence of massive particles elicit a set of important questions such as its implications for unitarity and the possibility of the mass inducing strong coupling effects. The correct treatment of these questions provided tools to understand the W-bosons and mass bounds on the Higgs  \cite{Lee:1977eg,Chanowitz:1978mv,Marciano:1989ns}.

In the context of the AdS/CFT correspondence  which is a conjectured duality  between ${\cal N}=4$ Super symmetric Yang-Mills (SYM) and string theory on $AdS_5\times S^5$ \cite{Maldacena:1997re}, the Higgs mechanism is well understood. It corresponds to taking the decoupling limit on the supergravity background describing two stacks of D3-branes keeping the distance between them  fixed, the distance between the two stacks is the dual to the mass of the W-boson in the field theory. The resulting supergravity background is explicitly known and has been discussed in the holographic setup \cite{Kraus:1998hv}.

Through a combination of modern unitarity methods and some string inspired approaches, a lot has been learned recently about the structure of  scattering amplitudes in general and in particular in ${\cal N}=4$ SYM \cite{Bern:2007dw}. It is fair to say that the spontaneously broken phase has received considerably less attention. Recently, Alday and Maldacena have proposed a prescription for computing some scattering amplitudes at strong coupling in the framework of the AdS/CFT correspondence. The prescription states that the color-ordered $n$-gluon MHV amplitude can be computed as \cite{Alday:2007hr} (see also reviews \cite{Alday:2008cg,Alday:2008yw})
\be
{\cal A}_n \sim \exp\left(-\frac{\sqrt{\lambda}}{2\pi}A(k_1, \ldots, k_n)\right),
\ee
where $A$ is the area of the minimal surface in the supergravity background that ends on a sequence of light-like segments on the boundary whose lengths is proportional to the momenta $k_i$.

The above prescription allows for a small modification that enables us to peek into the structure of the spontaneously broken phase of ${\cal N}=4$ supersymmetric Yang-Mills (SYM) with gauge group $SU(n_1+n_2) \to SU(n_1)\times SU(n_2)\times U(1)$. Namely, we consider the Alday-Maldacena prescription in the context of a supergravity background dual to the spontaneously broken phase of  ${\cal N}=4$. Although we are not able to solve the problem exactly, we consider various interesting approximations amenable to analytic work.

The paper is organized as follows. In section \ref{sec:AM} we review some technical aspects of the Alday-Maldacena prescription \cite{Alday:2007hr}. Section \ref{sec:Higgs} contains a discussion of the solution and the details of our evaluation of the amplitudes in specific kinematic regimes. Section \ref{sec:discussion} contains a discussion of our result from the field theoretic point of view and points out to some interesting open problems. We include two appendices, appendix \ref{appendix:perturbative} contains a discussion of the structure of the breaking of the gauge group and the appendix \ref{appendix:regularization} discusses an alternative way of regularizing some of the expressions presented in the main text.

\section{The Alday-Maldacena/Kruczenski solution}\label{sec:AM}

\subsection{General case}

We will be interested in classical world sheets embedded in spacetimes with
metrics of the following form:
\be
\label{background}
ds^2 =q^2(r)dx^\mu dx_\mu + p^2(r) dr^2.
\ee
Following \cite{Alday:2007hr}, we perform a T-duality and arrive at a metric of the form
\be
ds^2 =\frac{1}{q^2(r)}dy^\mu dy_\mu + p^2(r) dr^2.
\ee
This $T$-duality helps reformulate the scattering problem as a problem for a Wilson loop with simpler boundary conditions.
The induced metric on the worldsheet is then
\bea
ds_{WS}^2&=&\bigg[\frac{1}{q^2}\eta_{\m\n} \partial_1 y^\mu \partial_1 y^\nu +p^2 (\partial_1 r)^2 \bigg](d\s^1)^2
+\bigg[\frac{1}{q^2}\eta_{\m\n} \partial_2 y^\mu \partial_2 y^\nu +p^2 (\partial_2 r)^2 \bigg](d\s^2)^2 \nonumber \\
&+&2d\s^1 d\s^2\bigg[\frac{1}{q^2}\eta_{\m\n} \partial_1 y^\mu \partial_2 y^\nu+ p^2 \partial_1 r\partial_2 r\bigg].
\eea
We will assume configurations of the form
\be
\sigma^1=y^1, \qquad \s^2=y^2, \qquad y^0=y^0(y^1,y^2), \qquad r=r(y^1,y^2).
\ee
which leads to
\bea
\label{lagrangian}
S&=& \frac{1}{2\pi \alpha'}\int dy_1 dy_2 \frac{1}{q^2}\sqrt{1-(\partial_i y_0)^2 + p^2 q^2 (\partial_i r)^2
-p^2 q^2 (\partial_1 r \partial_2 y_0 - \partial_2 r \partial_1 y_0)^2 }. \nonumber
\eea
It is consistent to set the other $y^i$s to zero because they enter quadratically under the square root (and so their equations of motion are satisfied by $y^i=0$).

The specific cases of concern for us will be of the form:
\be
q^2=H(r)^{-\frac12}, \qquad p^2=H(r)^{\frac12},
\ee
specifically
\be
H(r)=\begin{cases} \frac{R^4 N_c}{r^4} & \text{for AdS} \\
R^4\left(\frac{n_1}{r^4}+\frac{n_2}{(r-a)^4}\right) & \text{for the Higgsed case} \end{cases}
\ee
Here, and throughout, we will use a non standard normalization $R^4=4\pi g_s l_s^4$, keeping the dependence on various $n_i$ explicit.

\subsection{The four-gluon amplitude}

Here we simply summarize the known woldsheet solution written by AM corresponding to the 4 point gluon scattering amplitude. This
solution was presented in \cite{Alday:2007hr} and was generated from the cusp solution of \cite{Kruczenski:2002fb}.  First, plugging in the AdS case into the above action, we find
\bea
S&=&\frac{1}{2\pi \alpha'}\int  dy_1  d y_2 \sqrt{g_{WS}} \\
&=& \frac{N_c^{\frac12}R^2}{2\pi \alpha'}\int dy_1  dy_2  \frac{1}{r^2}
\sqrt{1-(\partial_i y_0)^2+(\partial_i r)^2 -(\partial_1 r \partial_2 y_0 -
\partial_2 r \partial_1 y_0)^2 }. \nonumber
\eea
The solution for the generic case is
\bea
r&=&\frac{a}{\cosh u_1 \cosh u_2 + b\sinh u_1 \sinh u_2}, \quad
y_0 =\frac{a\sqrt{1+b^2}\sinh u_1 \sinh u_2}{\cosh u_1 \cosh u_2 + b\sinh u_1 \sinh u_2}, \nonumber \\
y_1&=&\frac{a\sinh u_1 \cosh u_2}{\cosh u_1 \cosh u_2 + b\sinh u_1 \sinh u_2}, \quad
y_2 =\frac{a\cosh u_1 \sinh u_2}{\cosh u_1 \cosh u_2 + b\sinh u_1 \sinh u_2}. \nonumber
\eea

To put the solution in the above form, we must invert the relations for $y_1$ and $y_2$.  This is accomplished by the change of coordinates
\bea
u_2={\rm arctanh} \left(\frac{a}{2 y_1 b}\left(1-\sqrt{1-4\frac{y_1 y_2 b}{a^2}}\right) \right) \nn \\
u_1={\rm arctanh} \left(\frac{a}{2 y_2 b}\left(1-\sqrt{1-4\frac{y_1 y_2 b}{a^2}}\right) \right) \nn \\
\eea
which gives
\bea
r &=& \Bigg(
\frac{\left(4y_1^2 b^2 -2 a^2+2 a \sqrt{a^2-4 b y_1 y_2}+4 b y_1 y_2\right)}{b y_1^2} \nn \\
&& \times\frac{\left(4y_2^2 b^2 -2 a^2+2 a \sqrt{a^2-4 b y_1 y_2}+4 b y_1 y_2\right)}{b y_2^2}\Bigg)^{1/2} \nn \\
&& \quad \times \frac{y_1 y_2}{2(a-\sqrt{a^2-4 b y_1 y_2})}\\
y_0  &=& \frac{1}{2b}\sqrt{(1+b^2)}\left(a-\sqrt{a^2-4 b y_1 y_2}\right).
\eea
Further, the relations to the Mandelstam variables is given by
\be
-s (2\pi)^2 = \frac{8a^2}{(1-b)^2}, \qquad -t(2\pi)^2 =\frac{8a^2}{(1+b)^2},
\ee
or inverting these relations we find
\be
a = \frac{\pi \sqrt{2}\sqrt{(-t)(-s)}}{(\sqrt{(-s)}+\sqrt{(-t)})},\quad  b = \frac{(\sqrt{-s}-\sqrt{-t})}{(\sqrt{-s}+\sqrt{-t})}.
\ee
%
\subsection{Evaluating the action, and the AM prescription}
The key physical information is encoded in the value of the action, and thus we need to compute it. The actions we wish to compute, as they stand, are
infinite. There are different ways of introducing a cut-off, and we outline a few below.

One may modify the solution somehow so that the boundary conditions are not met at $r=0$, but rather at $r=r_0$.  One can imagine doing this in two possible ways.  One may search for other solutions to the same action such that the boundary conditions are met at $r=r_0$, and then take a limit where $r_0\rightarrow 0$, and examine the divergences.  This appears to be the safest course of action, as one is always meeting the boundary conditions at every stage.  However, the above solutions may be hard to find, and so one may wish to consider an action which also depends on $r_0$, and so the action only collapses to the original action in the $r_0\rightarrow 0$ limit.  This has the utility of allowing for almost any function to be written down, however, one must be careful that the action converges to the desired action fast enough (this category of regulation includes the dimensional regularization used in \cite{Alday:2007hr}).

In the case of the ``wedge'' boundary condition (two lightlike lines), a solution with boundary conditions set at $r=r_0$ are now known exactly.  The new solution was presented in the last appendix of  \cite{Berkovits:2008ic}.
In appendix \ref{appendix:regularization} we consider the effects of regularizing with this solution.

A simpler approach, which was presented in \cite{Alday:2008cg}, simply takes the solution and cuts it off at $r=r_0$.  This has the utility of being simple, however the boundary conditions are only met in the limit that $r_0\rightarrow 0$, and only in a limiting sense.  In appendix \ref{appendix:regularization} we compare this type of regulation to that of fixing the boundary conditions at $r=r_0$ and then taking the limit as $r_0 \rightarrow 0$.

In this paper, we will mainly focus on the simple cutoff prescription.  We will not need to construct new solutions in this case, and so for our purposes is the easiest of the above regulations.

We now outline how to use the above classical string solution to determine a gauge invariant quantity in the gauge theory.  In \cite{Alday:2007hr} it was recognize that the factorization of planar (large $N_c$), ${\mathcal N}=4$ SYM scattering amplitudes offers a gauge invariant quantity one may be able to compute holographically.  The factorization of the amplitudes reads
\be
\mathcal{A}_n= A_{n,{\rm tree}} {\mathcal M}_n \label{factorization}
\ee
$A_{n,{\rm tree}}$ contains all lorentz and color indices.  This leaves ${\mathcal M}_n$ as a gauge invariant object that one may be able to compute using AdS/CFT.  In \cite{Alday:2007hr} (AM) they argue that the area of the classical string world sheet yields this piece of information.  More specifically, they argue that after performing a T-duality along the 4 flat spacetime directions $x_0... x_3$, one must compute the classical area of a worldsheet ending on $n$ lightlike line segments $p_i$.  These $n$ lightlike vectors are given by the $n$ lightlike momenta from the scattering amplitude one wishes to compute.  In \cite{Alday:2007hr}, they explicitly perform such a calculation for the 4-point amplitude, regulating the surface area using dimensional regularization, and in \cite{Alday:2008cg} via the simple cutoff prescription mentioned above.  The relation of this surface area to the quantity ${\mathcal M}_n$ is given by
\be
{\mathcal M}_n=e^{-\frac{\sqrt{\lambda}}{2 \pi}{\rm Area_2}}
\ee
where ${\rm Area}_2$ is the 2 dimensional area of the world sheet as given by the $p_i$.

Here, we will again be calculating the area of a classical world sheet, however, in a higgsed model.  As shown in  appendix A, the deep IR of the higgsed theory contains 3 copies of the $N=4$ gauge multiplets (with $SU(n_1), SU(n_2),$ and $U(1)$ gauge groups, respectively) after integrating out massive modes.  Hence, we expect the factorization of amplitudes to appear in the deep IR, keeping in mind that the different sectors do not interact.  Deep the UV, one expects that the vevs may be neglected, and so one again arrives at 1 copy of the $N=4$ gauge multiplets (with $SU(n_1+n_2)$ gauge group).  Again, one therefore expects factorization of the amplitudes in this limit.  We will assume here that factorization (\ref{factorization}), is always valid (although we emphasize that this is indeed an assumption), and so one may always define such an ${\mathcal M}_n$.  It would be interesting to see if this is indeed the case on the field theory side, and see if this follows from ${\mathcal N}=4$ supersymmetry, rather than from the full ${\mathcal N}=4$ superconformal symmetry. \footnote{This is seen most easily from the D3 brane picture: the number of killing spinors preserved by 2 parallel stacks is the same as one, hence 16 preserved supercharges (see \cite{Fayet:1978ig}).  The near horizon limit, however, does not introduce the conformal symmetry, thanks to the separation vector of the stacks.}

One final note is in order, and it helps elucidate some of the implications of assuming the factorization in (\ref{factorization}).  In (\ref{factorization}), $A_{\rm tree}$ is understood to have some values of the coupling constant $g$ in them.  Since this is a non conformal theory $g_1$ and $g_2$ (the couplings for $SU(n_1)$ and $SU(n_2)$ respectively) start as being different in the deep IR.  We will assume that the Lagrangian found in the appendix is appropriate, up to the couplings $g_i$ running with scale.  At the scale given by the masses of the $W$ bosons, these couplings should unify, and become the coupling constant for the $SU(n_1+n_2)$ gauge theory.  However, during the entire flow, the form of $A_{\rm tree}$ (up to the flow of $g$) remains the same when considering scattering within a given $SU(n_i)$ (i.e. restricting both incoming and outgoing particles).  This is because coupling to fields in the adjoint the other $SU(n_i)$ \footnote{henceforth we will refer to fields transforming in the adjoint of the $SU(n_1)$ as being ``in the $SU(n_1)$ sector''} is mediated by $W's$, and these must be produced in pairs, thanks to the extra $U(1)$.  Therefore, the only coupling between the $SU(n_1)$ and $SU(n_2)$ sectors occur at loop level.  However, because we assume that whole amplitude is proportional to the tree amplitude, we conclude that there is in fact no scattering between the $SU(n_1)$ sector and the $SU(n_2)$ sector (without producing $W$'s).  This is in fact the case for the pure $\mathcal{N}=4$ theory with unbroken $SU(n_1+n_2)$ gauge group in the planar limit, and is due to the factorization.  If one wishes to have both $SU(n_1)$ and $SU(n_2)$ fields in an interaction, one must in fact, also couple to at least one pair of $W$s during the interaction.  We do not allow loops, and so these $W$s are part of the asymptotic states (either incoming or outgoing) in the interaction.

Let us argue this from the standpoint of the world sheets near two stacks of D3 branes, which we label $p$ and $p'$.  First, we note that we are working at large $N$ and small $g_s$, so we wish to only consider disc diagrams.  Next, we consider inserting several vertex operators on this disc.  Let us denote vertex operators associated with $p-p$ strings as $V_X$, $p'-p'$ strings as $V_Y$ and $p'-p$ strings as $V_W$.  Because there is just one boundary, it is clear that a $V_X$ insertion cannot appear next to a $V_Y$ insertion, because the boundaries do not agree (further, the Chan-Paton factors do not agree).  Hence, there are no diagrams with only $V_X$ and $V_Y$ insertions, only insertions of the form $V_X V_X V_W V_Y V_Y V_W V_x \cdots$ are allowed.  The $V_W$ insertions are exactly the ``$W$'' fields mentioned above.  Hence, we expect this same behavior from the string side, at least in the strict $g_s\rightarrow 0$ limit.  One may need to be careful once considering the appropriate vertex operators in AdS, however.

Given the above assumptions, we will calculate the corrections to the amplitudes $\mathcal{M}_n$ holographically in the next section.

\section{Higgs Phase}\label{sec:Higgs}
It is of interest to understand how the mass of the higgs and the W-bosons can affect amplitudes.
We consider a simple model. Taking the decoupling limit of two stacks of D3 branes leads to a theory with a
Higgs branch where the vev of the Higgs field (or mass of the W-boson) is proportional to the distance between the stacks.

The supergravity solution has a metric
\bea
ds^2&=&H^{-1/2}dx_\mu dx^\mu + H^{1/2}dz_mdz^m, \nonumber \\
H&=&R^4\left(\frac{n_1}{\left(\overrightarrow{r}+\frac{\overrightarrow{a_0} n_2}{n_1+n_2}\right)^4}
+\frac{n_2}{\left(\overrightarrow{r}-\frac{\overrightarrow{a_0} n_1}{n_1+n_2}\right)^4}\right).
\eea
where $\overrightarrow{r}=(z^1,z^2,...,z^6)$ and $\overrightarrow{a_0}$ is a constant displacement vector.
Solving for gluon scattering in this background is challenging and in what follows we make a series of approximations.  First, however, we note that the above solution still has an SO(5) symmetry that leaves $\overrightarrow{a_0}$ fixed.  We use this vector to define a ``north pole'' in our $S^5$ coordinates.  The directions orthogonal to $\overrightarrow{a_0}$ appear in such a way that setting them to 0 satisfies the worldsheet equations of motion (they appear as functions of quadratic functions).  We therefore may consider only working with the world sheet at the "north pole" of these coordinates, and we will denote the coordinate along $\overrightarrow{a_0}$ as $r$.  In such a coordinate system, we find that
\be
H=R^4\left(\frac{n_1}{\left({r}+a_1\right)^4}
+\frac{n_2}{\left({r}-a_2\right)^4}\right)
\ee
such that the total separation is defined by $a_0=a_1+a_2$.

We will look at two approximations.  First, we will look at the case where the scale of interaction $a$ and radial cutoff $r_0$ are much bigger than the scale of separation of the two stacks.  Second, we will look a the case where $a_0$ is much larger than the other scales in the problem.  In these approximations, we will expand the action to the form
\be
S=S_0 + \epsilon S_1
\ee
where $S_0$ is an action exactly of the form considered by Alday and Maldacena.  To evaluate the correction to the total action, one simply needs to insert the 0th order solution into the corrected action.  Hence, in our approximation schemes with simple radial cutoff, we will not need to compute any new solutions.

Before doing this, however, a few simple observations are in order.  One should note that the T-duality keeps the harmonic functions intact, only affecting whether they come in a numerator, or denominator.  If one has a region of coordinates where one term or another dominates in the harmonic function, this is still true in the T dual coordinates.  For our simple case, near either stack of branes, we have an AdS throat.  However, in the T-dual coordinates, the throat becomes a boundary.
\begin{figure}[ht]
\centering
\includegraphics[width=.45\textwidth]{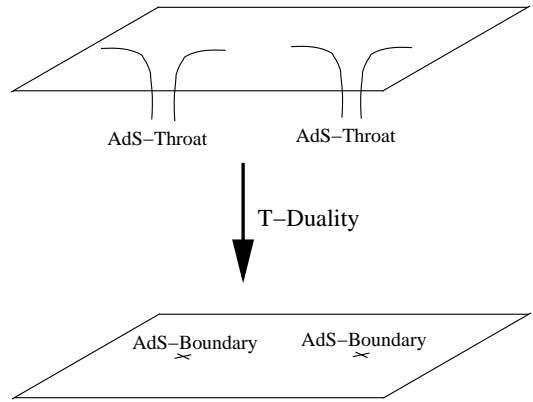}
\caption{The action of T-duality on the geometry of the background.}
\label{throattoboundary}
\end{figure}
These boundaries are the origin of the divergences in the values of our actions.  The the interpretation of these area divergences is that they are the IR divergences arising in the Feynman amplitude.

In our Higgsed theory we will have two sets of massless fields, namely two copies of $\mathcal{N}=4$, one for each throat.\footnote{In addition we have the extra $U(1)$ fields as well (the ``radion'' modes).  However, this U(1) may not be promoted to $U(2)$ by any addition of probe branes, as the $SU(n_i)$ may be promoted to an $SU(n_i+1)$.  This, then, does not allow for the same type of IR regulation that the two $SU(n_i)$ factors.  Hence, one cannot consider the IR regulation that we employ for scattering these fields, and so we only consider scattering fields in the adjoint of $SU(n_1)$ {\it or} in the adjoint of $SU(n_2)$.  It would be interesting to explore the bulk modes of this background, find the appropriate radion field (and superpartners), and try to match some of its properties to the Higgsed ${\mathcal{N}}=4$ theory.}  We wish to consider scattering of fields within a given $SU(n_i)$ sector, say the $SU(n_1)$ sector.  These $n$ point amplitudes are expected to have some IR divergences.  As mentioned above, we will consider regulating these IR divergences by a simple cutoff near the stack where the boundary conditions are imposed.

Note, from our stringy picture above, that when considering scattering only in the $SU(n_1)$ sector that the area divergences are only associated with the stack of $n_1$ branes, as the singularity structure of the metric near $r=-a_1$ is completely determined by this stack.\footnote{The divergences also depend on the boundary conditions imposed at the brane too.  For example we will be considering boundary conditions with cusps.  Other boundary conditions can also have divergences in the area of the worldsheet.  All of these divergences, however, depend on the infinite (spacelike) geodesic distance to the boundary of AdS, where the boundary conditions are imposed.}  This is because the boundary conditions in either T-dual frame must be satisfied at $r=-a_1$ when considering scattering within the $SU(n_1)$ sector.  We emphasize, however, that the above picture is valid only in the strong coupling and large $n_1, n_2$ regime.  Further, we also note that when the massive $W$ fields (with one $SU(n_1)$ index and one $SU(n_2)$ index) appear in external states we expect to have other types of divergences coming both throats, as the external $W$s stretch between the stacks of branes (onto each boundary in the T-dual frame).  One can see one such type of divergence quite easily.  Asymptotically, an external $W$ state will correspond to a string stretched between the two stacks following a straight timelike geodesic (because the $W$ is massive). This straight string configuration will have an area divergence associated with getting near to either $r=-a_1$ or $r=a_2$ because the geodesic is non null.  The procedure for imposing an IR cutoff, however, remains unclear, as one would presumably place 2 regulator branes near each boundary.  We speculate that one should place the 2 regulator branes such that they were each near their respective stack, but at the same ``potential'' value as measured by the harmonic function.  Then, one would relax the branes to the stacks, but keeping them always at the same ``potential'' value, thus only having one cutoff parameter given by the value of $H(r)$.

\subsection{$|s|,|t| \gg m_{IR}^2 \gg M_W^2$}

In this section, we will evaluate the action using mass parameter of the W as the perturbative parameter.  We expect this to be a good approximation when $a$ and the IR cutoff $r_0$ are both greater than the characteristic distance scale $a_0$ (this is the hierarchy in the subsection title, given that $s,t\sim a^2, m_{IR} \sim r_0, M_W\sim a_0$).  The sketch of this is shown in figure \ref{a0small}.

\begin{figure}[ht]
\centering
\includegraphics[width=.25\textwidth]{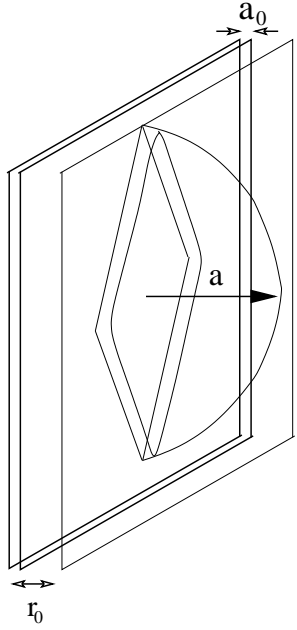}
\caption{The scales involved in the limit that $a_0$ is small, compared to the other scales in the problem, including the IR regulator $r_0$ and the scale of the interaction given by $a = \frac{\pi \sqrt{2}\sqrt{(-t)(-s)}}{(\sqrt{(-s)}+\sqrt{(-t)})}$, which determines roughly the depth in AdS into which the worldsheet falls.}
\label{a0small}
\end{figure}

For evaluating the expression, we use a ``center of mass coordinate'' such that the harmonic function
\be
H=R^4\left(\frac{n_1}{\left(\overrightarrow{r}+\frac{\overrightarrow{a_0} n_2}{n_1+n_2}\right)^4}
+\frac{n_2}{\left(\overrightarrow{r}-\frac{\overrightarrow{a_0} n_1}{n_1+n_2}\right)^4}\right)
\ee
where $\overrightarrow a_0$ is the displacement vector of the two stacks of branes.  Again, we will work at the ``north pole'' of the five sphere defined by this displacement vector.  In these coordinates, the harmonic function becomes
\be
H=R^4\left(\frac{n_1}{\left({r}+\frac{{a}_0 n_2}{n_1+n_2}\right)^4}
+\frac{n_2}{\left({r}-\frac{{a}_0 n_1}{n_1+n_2}\right)^4}\right)
\ee
where $a_0$ is the magnitude of the corresponding vector.  At $r$ larger than $a_0$, there is no linear term in $a_0$ by construction, and the expansion begins at second order.  Hence, we take the full action
\bea
S&=& \frac{R^2}{2\pi \alpha'}\int dy_1 dy_2 \sqrt{\left(\frac{n_1}{\left({r}+\frac{{a}_0n_2}{n_1+n_2}\right)^4}
+\frac{n_2}{\left({r}-\frac{{a}_0 n_1}{n_1+n_2}\right)^4}\right)} \nn \\
&&\quad \quad \times\sqrt{1-(\partial_i y_0)^2 +  (\partial_i r)^2
- (\partial_1 r \partial_2 y_0 - \partial_2 r \partial_1 y_0)^2 }.
\eea
and expand for $a_0\ll r$, and find the first two orders
\bea
S&=& \frac{R^2}{2\pi \alpha'}\int dy_1 dy_2 \frac{\sqrt{n_1+n_2}}{r^2}\sqrt{1-(\partial_i y_0)^2 +  (\partial_i r)^2
- (\partial_1 r \partial_2 y_0 - \partial_2 r \partial_1 y_0)^2 }. \nonumber \\
&&\!\!\!+\frac{R^2}{2\pi \alpha'}\int dy_1 dy_2 \frac{5\sqrt{n_1+n_2}n_1 n_2 a_0^2}{r^4 (n_1+n_2)^2}\sqrt{1-(\partial_i y_0)^2 +  (\partial_i r)^2
- (\partial_1 r \partial_2 y_0 - \partial_2 r \partial_1 y_0)^2 }. \nn \\
\eea
Note that in the above, the $n_1\rightarrow 0$ or $n_2\rightarrow 0$ limit yields no change to the action, as we should expect: if there isn't a second stack of branes, there is nothing new.

So our job as stated above is to evaluate the leading term in the action using the $0$th order solution.  We find that the general $s\neq t$ case difficult to analyze, and so we proceed by taking the simpler $s=t$ case.

\subsubsection{$s=t$}

As stated, we wish to evaluate the action
\bea
 && \Delta S= \\
 && \;\; \frac{R^2}{2\pi \alpha'}\int dy_1 dy_2 \frac{5\sqrt{n_1+n_2}n_1 n_2 a_0^2}{r^4(n_1+n_2)^2}\sqrt{1-(\partial_i y_0)^2 +  (\partial_i r)^2
- (\partial_1 r \partial_2 y_0 - \partial_2 r \partial_1 y_0)^2 } \nn
\eea
on the solution for $s=t$.

The $s=t$ case written in AM is given by
\be
r=\frac{\sqrt{(a^2-y_1^2)(a^2-y_2^2)}}{a}, \quad  y_0=\frac{y_1 y_2}{a}.
\ee
Defining a new set of variables
\be
y_1=\hat{y}_1 a, \quad y_2=\hat{y}_2 a
\ee
and defining a new set of dependent variables
\be
r=\hat{r} a, \quad y_0=\hat{y}_0 a
\ee
We find that the action scales to
\be
\Delta S(z,y_0)=\Delta \frac{1}{a^2}S(z,y_0)|_{a=1}
\ee
and so we will take the $a=1$ case, and simply scale it at the end.

Plugging the $a=1$ case into the action, we find
\be
\Delta S= \frac{R^2}{2 \pi \alpha'}\int_{-1}^1 dy_1 \int_{-1}^1 dy_2 \frac{5\sqrt{n_1+n_2}n_1 n_2a_0^2}{(n_1+n_2)^2(1-y_1^2)^2(1-y_2^2)^2}
\ee
This integral is divergent, and so we must regulate it.  We take the simple regulator that
\be
r=\epsilon= \rm{constant}=\sqrt{(1-y_1^2)(1-y_2^2)}
\ee
This restricts the bounds of integration in $y_1$ and $y_2$.
\be
y_1\in\left[-\sqrt{1-\epsilon^2},\sqrt{1-\epsilon^2}\right],\quad y_2\in\left[-\sqrt{\frac{1-y_1^2-\epsilon^2}{1-y_1^2}},\sqrt{\frac{1-y_1^2-\epsilon^2}{1-y_1^2}}\right]
\ee
Recall that the above $r=\epsilon$ is really $\hat{r}=\epsilon$.  This means that the cutoff will scale to $r=a \epsilon$.  We do not wish our regulator to be dependent on the scale of the collision, hence we take that $\epsilon=\frac{r_0}{a}$ where $r_0$ is now independent of $a$.  Recall also that we will be working in a regime where a great deal of the worldsheet extends beyond the $IR$ regulator.  Hence, we must have that $a\gg r_0$ and so $\epsilon$ is a unitless perturbative parameter: physically it is ``IR cutoff/scale of interaction.''

The only place that this cutoff appears is in the bounds of integration, and so we will expand these to next to leading order in $\epsilon$ such that we can trust the first 2 terms in this epsilon expansion.  This is taking the $\epsilon^0$ and $\epsilon^2$ terms in the bounds of integration.  We perform the $y_2$ integration, expand to next to leading order in $\epsilon$: one gets a divergence of the form $\epsilon^{-2}$ with logs, and an $\epsilon^0$ term with logs. We then perform the $y_2$ integration, and expand this to next order $\epsilon^2$ order as well.  Doing this, we find
\bea
\Delta S&=& \frac{R^2}{2 \pi \alpha'}\frac{\sqrt{n_1+n_2} n_1 n_2 a_0^2}{ a^2(n_1+n_2)^2}\Bigg[\frac{5}{4}\left(12\ln(2)-8\ln(\epsilon)+1\right)\epsilon^{-2} \nn \\
&&\quad +\frac{5}{48}\left(24\left(3\ln(2)-\ln(\epsilon)-\frac54\right)\left(\ln(2)-\ln(\epsilon)-\frac14\right)+\frac{57}{2}+2\pi^2\right)
\Bigg] \nn \\
\eea
plus terms that drop as $\epsilon^2$.
One may, if one wishes, reintroduce $a$ via $\epsilon=r_0/a$, and so explicitly see the IR cutoff $r_0$ and the scale of interaction $a\sim s=t$.

For comparison, we will also need the zeroth order action
\be
S= \frac{R^2}{2\pi \alpha'}\int dy_1 dy_2 \frac{\sqrt{n_1+n_2}}{r^2}\sqrt{1-(\partial_i y_0)^2 +  (\partial_i r)^2
- (\partial_1 r \partial_2 y_0 - \partial_2 r \partial_1 y_0)^2 }
\ee
and find that evaluated on the equations
\be
S_0=\frac{R^2}{2\pi \alpha'}\int dy_1 dy_2 \frac{\sqrt{n_1+n_2}}{(1-y_1^2)(1-y_2^2)}.
\ee
Using the same regulation technique as above, we find
\bea
S_0&=&\frac{R^2\sqrt{n_1+n_2}}{2\pi \alpha'}\Bigg[\fft{1}{12} \left(72 \ln(2)^2-3-96\ln(2)\ln(\epsilon)+24\ln(\epsilon)^2-2\pi^2\right) \nn \\
&&+\fft{1}{16}\left(-12\ln(2)+9+4\ln(\epsilon)\right)\epsilon^2 \Bigg].
\eea
Note that the most divergent term is
\be
\frac{R^2\sqrt{n_1+n_2}}{2\pi \alpha'}\frac{1}{12}24 \ln(\epsilon^2)\sim \frac{\sqrt{\lambda}}{2 \pi} \frac{1}{2}\ln(r_0^2)^2
\ee
which is what we needed to have the same answer as Alday in \cite{Alday:2008cg}.

Now a bit of numerology.  Note that the corrected action is more divergent by a power of $\epsilon^{-2}$.  This may be expected.  Given that we are dealing with a theory with massless propagators and then dealing with the mass of some of them perturbatively.  To be more explicit, we write out an arbitrary diagram with massive propagators.  We then expand these propagators for small mass $M$, i.e. $1/(p^2+M^2)\rightarrow 1/p^2-M^2/p^4$.  Therefore, when expanding to leading order, the integral will be 2 more powers divergent in $p$ around $p=0$ and hence go like $1/\epsilon^2$ where $\epsilon$ is the IR regulator.  This is exactly what we are seeing.

To fully see any effect, however, we would need to find the explicit solution near one of the stacks of branes, and so go into a region where $r \ll a_0$.  This would go beyond our approximation method here, where we have assumed $a_0 \ll r_0 < r$.

For completeness, we plug in $\epsilon=r_0/a=m_{IR}/a(s,t,u), a_0=M_W$ and
\be
a = \frac{\pi \sqrt{2}\sqrt{(-t)(-s)}}{(\sqrt{(-s)}+\sqrt{(-t)})},\quad  b = \frac{(\sqrt{-s}-\sqrt{-t})}{(\sqrt{-s}+\sqrt{-t})} \label{aeqn}
\ee
\footnote{Note that these substitutions do not match in mass dimension.  One may include relevant factors of $\alpha'$ if one wishes.  However, the action is unitless, and so all such factors divide out, and leave expressions exactly the same as doing the above substitutions.  For example, note that $a$ must be a length, where the RHS of (\ref{aeqn}) is a momentum, and so one must include an $\alpha'$ on the right to match units.} taking $s=t$, and find
\bea
&& \Delta S = \frac{R^2\sqrt{n_1+n_2}}{2\pi \alpha'}\frac{n_1 n_2 M_W^2 }{(n_1+n_2)^2}\fft54 \Bigg[
\frac{\left(12\ln(2)+1-4\ln\left[\frac{2 m_{IR}^2}{\pi^2(-s)}\right]\right)}{m_{IR}^2}  \\
&& \;\;-\fft16 \frac{\left[-6\left(6 \ln(2)-\ln\left(\frac{2 m_{IR}^2}{\pi^2(-s)}\right)-\fft52\right) \left(2 \ln(2)-\ln\left(\frac{2m_{IR}^2}{\pi^2(-s)}\right)-\fft12\right)+\fft{57}{2}+2 \pi^2\right]}{\pi^2 (-s)} \Bigg].\nn
\eea
One should note that while these terms look singular, the restrictions $M_W^2\ll m_{IR}^2\ll |s|$ confines the region of validity to where the above is small.

Similarly, we find that the original action is
\be
S_0= \frac{R^2\sqrt{n_1+n_2}}{2\pi \alpha'}
\left[\fft12\left(\ln\left(\frac{m_{IR}^2}{8\pi^2(-s)}\right)\right)^2-2\ln(2)^2-\fft14-\fft16 \pi^2\right]
\ee
which agrees exactly with Alday \cite{Alday:2008cg}.

\subsection{$|s|,|t| \ll M_W^2$: Effect of massive Ws in loops.}

\begin{figure}[ht]
\centering
\includegraphics[width=.45\textwidth]{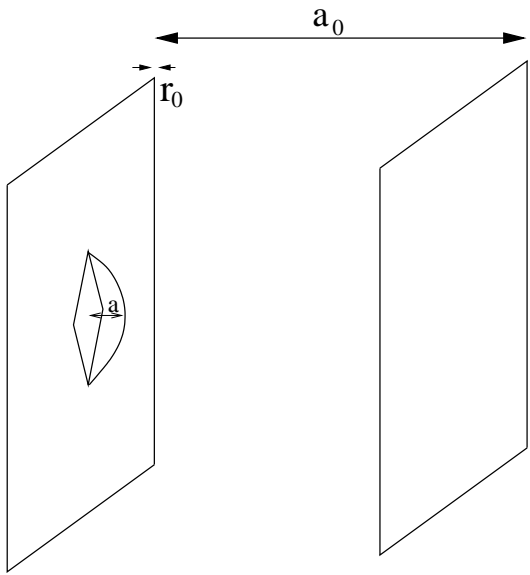}
\caption{The scales involved in the limit that $a_0$ is large, compared to the other scales in the problem, including the IR regulator $r_0$ and the scale of the interaction given by $a = \frac{\pi \sqrt{2}\sqrt{(-t)(-s)}}{(\sqrt{(-s)}+\sqrt{(-t)})}$, which determines roughly the depth in AdS into which the worldsheet falls.  Interestingly, the IR regulator does not play a role in the corrected action, as this piece is finite.  Therefore, the value of the corrected action is IR regulator independent.}
\label{a0big}
\end{figure}

There is another region of interest as well.  We will consider the large $W$ mass limit, and so we take $a_0\rightarrow \infty$.  In this limit, the above action becomes
\bea
&& S= \\
&&\;\;\frac{R^2\sqrt{n_1}}{2\pi \alpha'}\int dy_1 dy_2 \frac{1}{r^2}\left(1+\frac12\frac{r^4}{a_0^4}\frac{n_2}{n_1}\right)\sqrt{1-(\partial_i y_0)^2 +  (\partial_i r)^2
- (\partial_1 r \partial_2 y_0 - \partial_2 r \partial_1 y_0)^2 }. \nn
\eea
We expect the $r^4/a_0^4$ correction to be more convergent as it has extra powers of $r$ in the numerator.  We will in fact be able to evaluate this correction analytically, and do all relevant integrations.  This has the important quality that the answer does not depend at all on $r_0$, i.e. this information is IR regulator independent.  Further, we expect that higher order corrections continue to be convergent because the power series will only continue to have higher powers in $r$.

As mentioned above, to get the correction to the action to leading order, one simply needs to evaluate the old solution in the new action.  The solution in the $y_1,y_2$ coordinates is as follows
\bea
r &=& \Bigg(
\frac{\left(4y_1^2 b^2 -2 a^2+2 a \sqrt{a^2-4 b y_1 y_2}+4 b y_1 y_2\right)}{b y_1^2} \nn \\
&& \times\frac{\left(4y_2^2 b^2 -2 a^2+2 a \sqrt{a^2-4 b y_1 y_2}+4 b y_1 y_2\right)}{b y_2^2}\Bigg)^{1/2} \nn \\
&& \quad \times \frac{y_1 y_2}{2(a-\sqrt{a^2-4 b y_1 y_2})}\\
y_0  &=& \frac{1}{2b}\sqrt{(1+b^2)}\left(a-\sqrt{a^2-4 b y_1 y_2}\right).
\eea
Note that the $b=0$ limit is smooth.  We find that after plugging into the corrected Lagrangian, we can write it as
\bea
&& \Delta \mathcal L = \frac{R^2\sqrt{n_1}}{2\pi \alpha'}\times  \\
&&\frac{n_2}{n_1}\frac{ (-2y_1 +a(\mathcal{M}+1))(2y_1 +a(\mathcal{M}+1))(-2y_2 +a(\mathcal{M}+1))(2y_2 +a(\mathcal{M}+1))}{8 \mathcal{M} a_0^4 a^2(1+\mathcal{M}^2)}\nn
\eea
where we have defined
\be
\mathcal{M}\equiv \frac{\sqrt{a^2-4b y_1 y_2}}{a}.
\ee
Note that the branch of the square root appearing in the corrected action has been taken so that the integrand remained positive (which one can easily check at $y_1 = y_2 =0$).  The definition of $M$ is also well defined and unambiguous inside the range of integration (which we will discuss shortly).  To make the bounds of integration as easy as possible, we will further rotate to the coordinates $y_1'$ and $y_2'$ defined by
\be
y_1=\frac{1}{\sqrt{2}}\left(y_1' + y_2'\right)\quad y_2=\frac{1}{\sqrt{2}}\left(-y_1' + y_2'\right)
\ee
and then we will drop the primes.  The bounds of integration are given by where $z=0$, and these are determined on the four lines
\bea
&& y_2=\frac{(1-b)y_1-\sqrt{2}a}{(1+b)} \quad y_2=\frac{(1-b)y_1+\sqrt{2}a}{(1+b)}\nn \\
&& y_2=\frac{-(b-1)y_1-\sqrt{2}a}{(1+b)} \quad y_2=\frac{-(1-b)y_1+\sqrt{2}a}{(1+b)}
\eea
Further, we see that while inside of the diamond defined by these lines, $M$ is not imaginary.  An advantage of the coordinate change is now evident: One may compute the integral for the $y_1<0$ part of the diamond, and then double it (given the $y_i\rightarrow -y_i$ symmetry), and such a procedure is easiest in the above coordinates.  The mechanics of this are straightforward, but tedious.  We simply state the final result here
\be
\Delta S= \frac{R^2\sqrt{n_1}}{2\pi \alpha'}\frac{n_2 a^4 }{6 a_0^4 n_1}\frac{(1+b^2)\ln\left(\frac{(1+b)^2}{(1-b)^2}\right)-4b}{b^3}.
\ee
As a simple check, one may easily take a $b\rightarrow 0$ limit of the above, and compare it to the $b\rightarrow0$ limit of the original integral and see that they agree.  We may reexpress the above in terms of the Mandalstam variables $s$ and $t$ using
\be
a = \frac{\pi \sqrt{2}\sqrt{(-t)(-s)}}{(\sqrt{(-s)}+\sqrt{(-t)})},\quad  b = \frac{(\sqrt{-s}-\sqrt{-t})}{(\sqrt{-s}+\sqrt{-t})}
\ee
and make the replacement $a_0=M_W$.  Doing so, we find
\be
\Delta S= \frac{R^2\sqrt{n_1}}{2\pi \alpha'}\frac{n_2 \pi^4}{n_1 M_W^4}\left(
\fft43 \frac{(-t)^2(-s)^2((-s)+(-t))}{((-t)-(-s))^3}\ln\left(\frac{-t}{-s}\right)
-\fft83 \frac{(-t)^2(-s)^2}{((-t)-(-s))^2}
\right)
\ee
and so this action is symmetric under $s \leftrightarrow t$ as it should be.  Note also that the $s=t$ limit is smooth, as pointed out above (this is the $b=0$ limit).

To compare to the field theory, we note that
\be
\mathcal{A}=\exp(-S_{tot})=\exp(-S_{AM})\exp(\Delta S)=\exp(-S_{AM})\left(1-\Delta S\right)
\ee
so that the right hand factor may be interpreted as the loop diagrams with massive fields running in the loop (the subscript AM is Alday Maldacena).  Therefore, on the field theory side, we should express
\be
\mathcal{A}(s,t,M,r_0)=\mathcal{A}(s,t,M=\infty,r_0)\times \left(1+{\rm correction}\right).
\ee
We will qualitatively compare the correction we have found to field theory calculations in the next section.

\section{Discussion and outlook}\label{sec:discussion}
In this section we discuss the results of the previous sections largely by comparing them with similar results in field theory. Unfortunately, the field theory results are rather limited for our scope. For example, the general case of spontaneous breaking
$SU(n_1+n_2)\to SU(n_1)\times SU(n_2)\times U(1)$ has not been tackled in the literature. We have taken a modest step towards its perturbative understanding in appendix \ref{appendix:perturbative}. It would be interesting to pursue the field theoretic study of the effective action of the general case of spontaneously broken phase of ${\cal N}=4$ SYM with  gauge group $SU(n_1+n_2)$.

Some interesting questions have been raised in the special case of the above breaking: ${\cal N}=4$ SYM with gauge group $SU(2)$  spontaneously broken to $U(1)$. For example, the question of UV finiteness of the theory in this phase can be argued on general grounds but the concrete details of the cancelation are not spelt out in the literature. The precise structure of the low energy effective action was discussed in \cite{Buchbinder:2002tb}. Another important question in this context is whether the amplitudes are given only by box diagrams (in the perturbative regime). Although this is widely believed to be true, the full proof is lacking. The analysis of \cite{Buchbinder:2002tb} provides strong evidence that the answer is in the affirmative. Explicit computation of scattering amplitudes is another are where results are scarce. For example, explicit amplitudes can be found in \cite{Kubo:1984hm,Schabinger:2008ah}. Interestingly, the work of \cite{Schabinger:2008ah} provides further evidence to the hypothesis that in the case of ${\cal N}=4$ SYM with gauge group  $SU(2)$ spontaneously broken to $U(1)$ the scattering amplitudes, including those with external massive states are box type.

According to the Alday-Maldacena prescription \cite{Alday:2007hr}, our calculation should be interpreted as ${\cal A}_4=A_{tree}{\cal M}_4$. Since our ${\cal M}_4$ contains only terms of the form $1/M_W^4$ we do not foresee terms of the form $M_W^6$ in the expansion that could potentially indicate triangle contribution. Moreover, since the expression for ${\cal M}_4$  contains only singularities that can be induced at one-loop some hope remains that the full amplitude allows for a type of exponentiation Ansatz similar to generalizations of the BDS one \cite{Bern:2005iz}.   We recall the heart of the BDS ansatz for the scattering cross section, and it reads
\be
A=A_{tree}\times e^{f(\lambda)\frac{\mbox{1-loop}}{\mbox{tree-level}}}.
\ee
Let us take that the form of the factorization that we have is exactly the same.  Hence, the 1 loop result contain 2 pieces: those with only massless fields running in the loops, and ones with massive fields running in the loops.  We further only consider box diagrams, as the preceding paragraph suggests.  Also, recall that massive fields can only be created or destroyed in pairs.  With these pieces of information, we can conclude that the 1 loop amplitudes have only 2 pieces: a box diagram with massless fields running in the loop (which comes from the theory with the massive fields integrated out), and a box diagram with massive fields running in the loop.  Hence, the only new contribution comes from the new box diagram.  In what follows, we will look at this box integral in various limits, and be able to read off directly the part we need to compare to our corrected actions.  Implicit in this is the assumption that the running of the coupling $\lambda$ (or equally $g$) does not contribute to the quantity we are calculating, although this too appears as an overall contribution.

With this preamble we proceed to discuss our results. First we notice that the structure of singularities that we found is of the 2 following forms:
In the case of $s,t>> M_{IR}^2>> M_{W}^2$, we found
\be
\frac{1}{M_{IR}^2}, \quad \frac{\ln M_{IR}}{M_{IR}^2}, \quad \ln^2M_{IR}, \quad \ln M_{IR};
\ee
In the other case $|s|,|t| \ll M_W^2$, we found
\be
\frac{1}{M_W^4},
\ee
and we have a completely finite answer.  We wish to compare these qualitatively with the field theory side.  Because we are taking the same form of the factorization given above, we can compare directly the 1 loop result to our correction to the action (as both are exponentiated, and directly added to the original result).

Our regularization scheme is not dimensional regularization, the preferred one from the field theory point of view. However the nature of the divergences should be the same, such that the difference in physical quantities using different regulators is finite.  We can safely conclude that this structure of singularities is compatible with box integrals in dimensional regularization. The general structure of box diagrams in this scheme can be found in \cite{Ellis:2007qk}
and for the case of all propagators massive in \cite{Denner:1991qq}, see also \cite{Bern:1995db}. For example, the MHV one-loop amplitude for four gluons was obtained in \cite{Schabinger:2008ah}
\bea
&&\kern-1em {\cal M} (k_1^+,k_2^+,k_3^-, k_4^-)=8g^2 s^2 \left(I_0^{(4)}(s,t)+ I_0^{(4)}(s,u)+I_0^{(4)}(t,u)\right), \\
&& \kern-1em I_0^{(4)}(s,t)=\int \frac{d^4l}{(2\pi)^4}\frac{1}{(l^2+m^2)((l-k_1)^2+m^2)((l-k_1-k_2)^2+m^2)((l+k_4)^2+m^2)}. \nonumber
\eea
with the standard Mandelstam variables
\be
s=(k_1+k_2)^2, \quad t=(k_1+k_4)^2, \quad u=(k_1+k_3)^2.
\ee

Let us first compare to the large mass limit $|s|,|t| \ll M_W^2$.  One may easily rescale the values of the momenta above by $m$ and find
\bea
&& \kern-1em I_0^{(4)}(s,t)=\frac{1}{m^4}\int \frac{d^4\hat{l}}{(2\pi)^4}\frac{1}{(\hat{l}^2+1)((\hat{l}-\hat{k}_1)^2+1)((\hat{l}-\hat{k}_1-\hat{k}_2)^2+1)((\hat{l}+\hat{k}_4)^2+1)}. \nonumber
\eea
where we write the unitless momenta $\hat{l}=\frac{l}{m}$.  Identifying $m=M_W$ above, we see that the first correction does in fact go as $\frac{1}{M_W^4}$, with some unitless (and finite!) coefficient depending on kinematic information contained in the $k_i$.  This is qualitatively similar to the answer that we obtained in the strong coupling regime ($\frac{{\rm finite}}{M_W^4}$).  It would be interesting to try to expand this and compare to what we have more quantitatively, and help to decide whether accounting for the running in the $g_i$ is included in this overall factor.

We may also wish to look at the other regime, where  $s,t>> M_{IR}^2>> M_{W}^2$.  In such a situation, one is taking that the mass $M_W$ is small, and so the box integral above has new IR divergences (if one sets $m=0$ for example).  This means that we have to set a new IR regulator $M_{IR}$ to cut this off.  Doing so will give divergences that approximate the fields as massless first, i.e. in the theory with the full $SU(n_1+n_2)$ restored, and then have an $M_w^2/p^2$ type of correction.  This second type of term will give exactly the correction we have computed, and with the same kind of new divergences we see.  The new divergences will come with 2 extra powers of $M_{IR}$ in the denominator.  At the very least, we see that our expression is compatible with the structure of infrared singularities at loop $L$ this being of the form ${\cal M}_4^{(L)}\sim 1/M_{IR}^{2L}$ (after we exponentiate this result).  Unfortunately, we have been unable to find a convenient expansion of this part of the box integral for direct comparison.

It is worth mentioning that the analysis discussed here shares some interesting aspects with more phenomenologically relevant calculations like $\gamma \gamma \to \gamma \gamma$  in the standard model discussed in \cite{Jikia:1993tc,Gounaris:1999gh} where W-bosons or heavy quarks are allowed to run in loops. Similarly, in the limit of
large top quark mass the amplitudes for Higgs plus gluons simplify tremendously as first shown in \cite{Dawson:1991au}. More recently, the computation of Higgs boson plus an arbitrary number of partons \cite{Dixon:2004za}, confirmed the persistence of such relatively simple structure. Considering quarks in the context of the AdS/CFT also indicates a rather simple structure for the amplitudes \cite{Komargodski:2007er,McGreevy:2007kt}. We hope that our analysis will help uncover simple structure in the case of spontaneously broken phases.

\section*{Acknowledgements}
B.B. wishes to thank Jacob Sonnenschein and Luca Mazzucato for useful discussions.
The work of B.B. has been supported in part by NSERC of Canada,  and in part by the Israel Science Foundation under a grant (DIP H.52) of the German Israel Project Cooperation D.I.P. and by the European
Network MRTN-CT-2004-512194. L.A.P.Z. is  partially supported by Department of Energy under
grant DE-FG02-95ER40899 to the University of Michigan.

\appendix

\section{Appendix A: $SU(n_1+n_2)\rightarrow SU(n_1)\times SU(n_2)\times U(1)$}\label{appendix:perturbative}

In this section, we will be concerned with decomposing fields in the adjoint of $SU(n_1+n_2)$ into representations of the subgroup $SU(n_1)\times SU(n_2)\times U(1)$ to facilitate examination of the Higgsing of the $\mathcal N=4$ theory.  First, we note that the generators of a general $SU(N)$ can be written as diagonal elements
\be
\tau^{I=p}=\frac{1}{\sqrt{p^2+p}}\rm{diag}(1,1,1, ...1, -p,0,0,0)\quad p\in(1...N-1)
\ee
and $p$ ones appear before the final $-p$.  The remaining $N^2-1-(N-1)=N(N-1)$ generators are the off diagonal elements. There are $N(N-1)/2$ entries in the upper right hand triangle of the $N\times N$ matrix.  We take one entry to be $\frac{1}{\sqrt{2}}$, or $\frac{i}{\sqrt{2}}$, and the rest zero (therefore there are $N(N-1)$ of these).  The lower left triangle is determined by the property $(\tau^a)^\dagger=\tau^a$.  Such matrices clearly form a basis for the set of traceless Hermetian matrices that define the Lie algebra of $SU(N)$.

Such a basis also satisfies
\be
\Tr(\tau^I \tau^J)=\delta^{IJ}
\ee
and we define the structure constants via
\be
[\tau^I,\tau^J]=if^{IJK}\tau^{K}.
\ee
\footnote{If one wants to satisfy the usual $\Tr(\tau^I \tau^J)=\frac12 \delta^{IJ}$, one simply scales $\tau^I=\alpha T^I, f^{IJK}=\alpha F^{IJK}, g=G\alpha^{-1}$ with $\alpha=\sqrt{2}^{-1}$.  The action is invariant under such a scaling (recall that in component notation, no generators appear, and further $gf^{abc}$ always comes together: this is because we only deal with one representation of the gauge group).}  For the off diagonal components, it is often convenient to use a ``$\pm$'' basis, where one takes the two off diagonal elements with entries in the same spot (different only because of the $i$) and constructing $\tau^q+i\tau^{q+1}=\tau^{+q}$ ($q$ odd).  The hermeitain conjugate we call  $\tau^{-q}$.  These matrices can be chosen to be real, and have therefore one single non 0 entry that is 1.  We will only use this notation when we Higgs the theory.

For breaking $SU(n_1+n_2)\rightarrow SU(n_1)\times SU(n_2) \times U(1)$ the following generator is important
\be
\tau^0={\rm diag}\left(\frac{1}{n_1 e},\cdots,\frac{-1}{n_2e},\cdots\right).
\ee
Above, the entry $\frac{1}{n_1 e}$ happens $n_1$ times, the entry $\frac{-1}{n_2 e}$ happens $n_2$ times, and $e$ is defined to be
\be
e\equiv \frac{\sqrt{n_1+n_2}}{\sqrt{n_1 n_2}}.
\ee
Again, $\Tr((\tau^0)^2)=1$ because of
\be
\frac{1}{n_1}+\frac{1}{n_2}=e^2.
\ee

Fields in the adjoint representation of $SU(n_1+n_2)$ we represent in the following way
\be
\hat{\Sigma}^I\tau^I \equiv \hat{\Sigma}=
\begin{pmatrix}\sig1+\frac{\sig0}{n_1 e} & \omegs{+} \\
\omegs{-} & \sig2-\frac{\sig0}{n_2 e}
\end{pmatrix}
\ee
where $\sig1=\sig1^a \tau^a$ and $\tau^a$ are the generators of $SU(n_1)$, and similarly for $\sig2$.  $\sig0$ is understood to multiply the appropriate identity matrix.  Further, $\omegs{+}$ is an $n_1 \times n_2$ matrix, and $\omegs{-}$ is an $n_2 \times n_1$ matrix.  Clearly what we have done is taken the off diagonal elements in the $n_1 \times n_2$ block and reexpressed these in an appropriate $\pm$ basis, leaving the other off diagonal elements alone.  Looking ahead, the Higgsing effect will be giving the $\sig0$ part of some scalar field a vev (i.e. $<\hat{\Sigma}>=a\tau^0$).  This has the appropriate symmetry properties: the upper left $SU(n_1)$ and lower right $SU(n_2)$ matrices commute with this generator.  The unbroken $U(1)$ is generated by $\tau^0$ itself.

First, we decompose the gauge field as above, and denote
\bea
\hat{A_\mu}^I\tau^I \equiv \hat{A_\mu}=
\begin{pmatrix}\gau1_\mu+\frac{\gau0_\mu}{n_1 e} & \Ws{+}_\mu \\
\Ws{-}_\mu & \gau2_\mu-\frac{\gau0_\mu}{n_2 e}
\end{pmatrix}.
\eea
Note that we are using a real gauge field, hence we have the restriction that $(\hat{A}_\mu^I \tau^I)^\dagger=\hat{A}_\mu^I \tau^I$.  This translates to the following restriction $(\Ws{+}_\mu)^{\dagger}=\Ws{-}_\mu$.

We calculate the gauge covariant derivative acting on $\hat{\Sigma}$ \footnote{The gauge transformations are $\hat{\Sigma}'=\hat{\sigma}-i[\Lambda,\hat{\Sigma}]$ and $g \hat{A}'_\mu= g \hat{A}_\mu + \pa_\mu\Lambda+ig[\Lambda,\hat{A}_\mu]$ to linear order in $\Lambda$.}
\bea
&& \kern-2em D_\mu \hat{\Sigma} \equiv \pa_\mu -ig[\hat{A}_\mu,\hat{\Sigma}] =  \label{genCovD}\\
&&\kern-2em \begin{pmatrix}
D_\mu \sig1-ig\left(\Ws{+}_\mu \omegs{-}-\omegs{+}\Ws{-}_\mu\right)+\frac{\pa_\mu \sig0}{n_1 e} & D_\mu\omegs{+}-ig\left(\Ws{+}_\mu \sig2-\sig1 \Ws{+}_\mu-e\Ws{+}_\mu \sig0\right) \\
D_\mu \omegs{-}-ig\left(\Ws{-}_\mu \sig1-\sig2 \Ws{-}_\mu+e\Ws{-}_\mu\right) & D_\mu \sig2-ig\left(\Ws{-}_\mu \omegs{+}-\omegs{-}\Ws{+}_\mu\right)-\frac{\pa_\mu \sig0}{n_2 e}
\end{pmatrix}\nn
\eea
where the remaining gauge covariant derivatives are defined as follows
\bea
D_\mu \sig1&=&\partial_\mu \sig1 -ig[\gau1_\mu,\sig1] \\
D_\mu \sig2&=&\partial_\mu \sig2 -ig[\gau1_\mu,\sig2] \\
D_\mu \omegs{-}&=&\partial_\mu \omegs{-} -ig\left(\gau2_\mu \omegs{-}-\omegs{-} \gau1_\mu-e\gau0_\mu\omegs{-}\right)\\
D_\mu \omegs{+}&=&\partial_\mu \omegs{+} -ig\left(\gau1_\mu \omegs{+}-\omegs{+} \gau2_\mu+e\gau0_\mu\omegs{+}\right).
\eea
This is intuitively obvious: the traceless upper left block transforms as an adjoint of $SU(n_1)$, $\omegs{+}$ as a fundamental under $SU(n_1)$ and an antifundamental under $SU(n_2)$ and charge $+ge$ under the $U(1)$ generated by $\tau^0$.  Similar comments hold for $\omegs{-}$ and the traceless lower right part of the field $\hat{\Sigma}$.  Below we show the chart of charges for the various fields under $SU(n_1)\times SU(n_2)\times U(1)$.
\begin{center}
\begin{tabular}{c || c | c | c}
        & $SU(n_1)$ & $SU(n_2)$ & $U(1)$ \\
        \hline
$\sig1$ & Adj & 1 & 0 \\
$\sig2$ & 1 & Adj & 0 \\
$\omegs{+}$ & $\square$ & $\bar{\square}$ & $+ge$ \\
$\omegs{-}$ & $\bar{\square}$ & $\square$ & $-ge$ \\
$\sig0$ & 1 & 1 & 0
\end{tabular}
\end{center}
In all that follows, any covariant derivatives will be as the above, with the pre-subscript and $\pm$ superscripts denoting the charges.

The action we wish to reexpress has the following field content: first we have $6(N^2-1)$ complex scalars denoted
\bea
\hat{\Phi}^I_{ij}=-\hat{\Phi}^I_{ji}\quad i,j \in {1,2,3,4}
\eea
with the following restriction
\bea
(\hat{\Phi}^I_{ij})^{*}=\frac12 \epsilon^{i j k l} \hat{\Phi}^I_{k l} \equiv \hat{\Phi}^{I i j}.\label{sdeq}
\eea
This means we really have $3(N^2-1)$ complex scalars, i.e. $6(N^2-1)$ real scalars.  Each field $\hat{\Phi}^I_{ij}$ transforms as a {\bf 6} of SU(4).  Next, we have $4(N^2-1)$ dirac spinors which we project down to $4(N^2-1)$ Weyl spinors using $L=\frac{1+\gamma_5}{2}$ and $R=1-L$:
\be
L(\hat{\chi}^I)^{i}.
\ee
$(\hat\chi^I)^i$ is a {\bf 4} of SU(4).  We also have $(N^2-1)$ gauge fields, already denoted above as
\be
\hat{A}^{I}_\mu.
\ee
We decompose these fields in the following way
\bea
\hat{A_\mu}^I\tau^I \equiv \hat{A_\mu}&&=
\begin{pmatrix}\gau1_\mu+\frac{\gau0_\mu}{n_1 e} & \Ws{+}_\mu \\
\Ws{-}_\mu & \gau2_\mu-\frac{\gau0_\mu}{n_2 e}
\end{pmatrix}\\
\hat{\chi}^{I i} \tau^I \equiv \hat{\chi}^i&&=
\begin{pmatrix}\ch1^i+\frac{\ch0^i}{n_1 e} & \xis{+i} \\
\xis{-i} & \ch2^i-\frac{\ch0^i}{n_2 e}
\end{pmatrix}\\
\hat{\Phi}^{I}_{ij} \tau^I \equiv \hat{\Phi}_{ij}&&=
\begin{pmatrix}\pp1_{ij}+\frac{\pp0_{ij}}{n_1 e} & \oms{+}_{ij} \\
\oms{-}_{ij} & \ch2-\frac{\ch0}{n_2 e}
\end{pmatrix}.
\eea
We also define the following conventions
\bea
\overline{(\ch1^i)}&&\equiv \chb1_i \nn \\
\overline{(\ch2^i)}&&\equiv \chb2_i \nn \\
\overline{(\xis{+i})}&&\equiv \xibs{-}_i \nn \\
\overline{(\xis{-i})}&&\equiv \xibs{+}_i
\eea
where the lowered $SU(4)$ index indicates that it transforms as a $\bar{{\bf 4}}$ of $SU(4)$.
Finally, because of constraint (\ref{sdeq}), we have that
\bea
(\pp1_{ij})^{\dagger}&=&\frac12 \epsilon^{ijkl} \pp1_{kl}\equiv \pp1^{ij} \nn \\
(\pp2_{ij})^{\dagger}&=&\frac12 \epsilon^{ijkl} \pp2_{kl}\equiv \pp2^{ij} \nn \\
(\oms{+}_{ij})^{\dagger}&=&\frac12 \epsilon^{ijkl}\oms{-}_{ij} \equiv \oms{-ij} \\
(\oms{-}_{ij})^{\dagger}&=&\frac12 \epsilon^{ijkl}\oms{+}_{ij} \equiv \oms{+ij}
\eea
where the $\equiv$ is meant as the definition of the fields with the SU(4) indices up.

Using the above definitions, we write out the following terms.  First, defining
\be
\hat{F}_{\mu \nu}=2\pa_{[\mu \phantom]}A_{\phantom[ \nu]}-ig[\hat{A}_\mu,\hat{A}_\nu]
\ee
we find
\be
\hat{F}_{\mu \nu}=\begin{pmatrix}
\gauf1_{\mu \nu}+\frac{1}{n_1 e}\gauf0_{\mu \nu} -ig 2 \Ws{+}_{[\mu \phantom]}\Ws{-}_{\phantom[ \nu]} & 2 D_{[\mu \phantom]}\Ws{+}_{\phantom[ \nu]} \\
2 D_{[\mu \phantom]}\Ws{+}_{\phantom[ \nu]} & \gauf2_{\mu \nu}-\frac{1}{n_2 e}\gauf0_{\mu \nu} -ig 2 \Ws{-}_{[\mu \phantom]}\Ws{+}_{\phantom[ \nu]}
\end{pmatrix}
\ee
where we have defined $\gauf1$, $\gauf2$, and $\gauf0$ as above (of course with the commutator vanishing for $\gauf0$).  This gives quite trivially
\bea
&&-\frac14\Tr\left(\hat{F}_{\mu \nu}\hat{F}^{\mu \nu}\right)=\nn \\
&&-\frac14\tr1\Bigg[\left(\gauf1_{\mu \nu}+\frac{\gauf0_{\mu \nu}}{n_1 e} -ig 2 \Ws{+}_{[\mu \phantom]}\Ws{-}_{\phantom[ \nu]}\right)\left(\gauf1^{\mu \nu}+\frac{\gauf0^{\mu \nu}}{n_1 e} -ig 2 \Ws{+[\mu \phantom]}\Ws{-\phantom[ \nu]}\right) \nn \\
&&\qquad \qquad +2D_{[\mu \phantom]}\Ws{+}_{\phantom[ \nu]}2D^{[\mu \phantom]}\Ws{-\phantom[ \nu]}\Bigg] \nn \\
&&-\frac14\tr2\Bigg[\left(\gauf2_{\mu \nu}-\frac{\gauf0_{\mu \nu}}{n_2 e} -ig 2 \Ws{-}_{[\mu \phantom]}\Ws{+}_{\phantom[ \nu]}\right)\left(\gauf2^{\mu \nu}-\frac{\gauf0^{\mu \nu}}{n_2 e} -ig 2 \Ws{-[\mu \phantom]}\Ws{+\phantom[ \nu]}\right) \nn \\
&&\qquad \qquad +2D_{[\mu \phantom]}\Ws{-}_{\phantom[ \nu]}2D^{[\mu \phantom]}\Ws{+\phantom[ \nu]}\Bigg].
\eea
Above, we have explicitly written out $\Tr=\tr1+\tr2$ to emphasize which kind of indices are being traced over, even though this is evident from the term being traced.

Above, there are several terms that are zero.  For example, traces of the form $\tr1(\gauf1 \gauf0)$ vanish, as the trace of the $SU(n_1)$ matrices are zero.
Further, because of the cyclicity of the trace, traces of the form $\tr1(\Ws{+} \Ws{-})=\tr2(\Ws{-} \Ws{+})$,
which allows us to combine certain terms.  Further, one should note that the total term
$-\frac14\left(\tr1(\gauf0_{\mu \nu}\gauf0^{\mu \nu})/(en_1)^2+\tr2(\gauf0_{\mu \nu}\gauf0^{\mu \nu})/(en_2)^2\right)$
$=-\frac14\gauf0_{\mu \nu}\gauf0^{\mu \nu}$.  This normalized coefficient is just the statement that $\Tr((\tau^0)^2)=1$.
One can use these relationships to rewrite the above as
\bea
-&& \frac14\Tr(\hat{F}_{\mu \nu}\hat{F}^{\mu \nu})= \nn \\
&&\quad -\frac14\tr1(\gauf1_{\mu \nu}\gauf1^{\mu \nu})-\frac14\tr2(\gauf2_{\mu \nu}\gauf2^{\mu \nu})-\frac14\gauf0_{\mu \nu}\gauf0^{\mu \nu} \nn \\
&&\quad-\frac14\left[\tr1\left(2D_{[\mu\phantom]}\Ws{+}_{\phantom[\nu]}2D^{[\mu\phantom]}\Ws{-\phantom[\nu]}\right)
+\tr2\left(2D_{[\mu\phantom]}\Ws{-}_{\phantom[\nu]}2D^{[\mu\phantom]}\Ws{+\phantom[\nu]}\right)\right] \nn \\
&&\quad+ig\tr1\left(\gauf1^{\mu \nu}\Ws{+}_\mu \Ws{-}_\nu\right)+ig\tr2\left(\gauf2^{\mu \nu}\Ws{-}_\mu \Ws{+}_\nu\right) \\
&&\quad+ige\frac12\gauf0^{\mu\nu}\left[\tr1(\Ws{+}_\mu \Ws{-}_\nu)-\tr2(\Ws{-}_\mu\Ws{+}_\nu)\right] \nn \\
&&\quad-g^2\tr1\left(\Ws{+}_{[\mu \phantom]}\Ws{-}_{\phantom[ \nu]}\Ws{+[\mu \phantom]}\Ws{-\phantom[ \nu]}\right)-g^2\tr2\left(\Ws{-}_{[\mu \phantom]}\Ws{+}_{\phantom[ \nu]}\Ws{-[\mu \phantom]}\Ws{+\phantom[ \nu]}\right). \nn
\eea
The terms on the third and fifth lines could be combined further into single terms using the cyclicity of the trace.  However, we leave it in the above presentation to exhibit the symmetry $+\leftrightarrow -$, $1\leftrightarrow 2$ (referring to the $SU(n_i)$ factor), $e\rightarrow -e$, as it must be from the onset of the problem.  In this expanded form, it is clear that when the $W$'s are given a mass from a vev, and integrated out, the remaining theory will have the promised gauge symmetry of $SU(n_1)\times SU(n_2)\times U(1)$.

Similarly, one calculates
\bea
&&i \Tr\left(\bar{\hat{\chi}}_i\gamma^\mu D_\mu L \chi^i \right)= \nn \\
&& i\tr1\Bigg[\left(\chb1_i+\frac{\chb0_i}{n_1e}\right)\gamma^{\mu}L\left(D_{\mu}\ch1^i-ig\left(\Ws{+}_\mu \xis{-i}-\xis{+i}\Ws{-}_\mu\right)+\frac{\pa_{\mu}\ch0^i}{n_1 e}\right)\Bigg] \nn \\
&&+i\tr1\Bigg[\xibs{+}_i\gamma^{\mu}L\left(D_{\mu}\xis{-i}-ig\left(\Ws{-}_\mu \ch1^i-\ch2^i\Ws{-}_\mu+e\Ws{-}_\mu \ch0^i\right)\right)\Bigg]  \nn \\
&& i\tr2\Bigg[\left(\chb2_i-\frac{\chb0_i}{n_2e}\right)\gamma^{\mu}L\left(D_{\mu}\ch2^i-ig\left(\Ws{-}_\mu F
\xis{+i}-\xis{-i}\Ws{+}_\mu\right)-\frac{\pa_{\mu}\ch0^i}{n_2 e}\right)\Bigg] \nn \\
&&+i\tr2\Bigg[\xibs{-}_i\gamma^{\mu}L\left(D_{\mu}\xis{+i}-ig\left(\Ws{+}_\mu \ch2^i-\ch1^i\Ws{+}_\mu-e\Ws{+}_\mu \ch0^i\right)\right)\Bigg]
\eea
or again rearranging terms,
\bea
&&=i\tr1\left(\chb1_i\gamma^\mu D_\mu L\ch1^i\right)+i\tr2\left(\chb2_i\gamma^\mu D_\mu L\ch2^i\right) + i\chb0_i\gamma^\mu \pa_\mu L\ch0^i \nn\\
&&+g\tr1\left(\chb1_i\gamma^\mu[\Ws{+}_\mu L\xis{-i}-L\xis{+i}\Ws{-}_\mu]\right)+g\tr2\left(\chb2_i\gamma^\mu[\Ws{-}_\mu L\xis{+i}-L\xis{-i}\Ws{+}_\mu]\right) \nn \\
&&+\frac12ge\chb0_i \gamma^{\mu}\left[\tr1\bigg(\Ws{+}_\mu L\xis{-i}-L\xis{+i}\Ws{-}_\mu\bigg)-\tr2\bigg(\Ws{-}_\mu L\xis{+i}-L\xis{-i}\Ws{+}_\mu\bigg)\right] \nn \\
&&+i\tr1\left(\xibs{+}_i\gamma^\mu \left[D_\mu L\xis{-i}-ig(\Ws{-}_\mu L\ch1^i-L\ch2^i\Ws{-}_\mu+e\Ws{-}_\mu L\ch0^i)\right]\right) \nn \\
&&\quad \quad \quad +i\tr2\left(\xibs{-}_i\gamma^\mu\left[D_\mu L\xis{+i}-ig(\Ws{+}_\mu L\ch2^i-L\ch1^i\Ws{+}_\mu-e\Ws{+}_\mu L\ch0^i)\right]\right).
\eea
For the scalars, one calculates
\bea
&&\frac12\Tr\left(D_\mu \hat{\Phi}_{ij} D^\mu \hat{\Phi}^{ij}\right) = \nn \\
&&\frac12\tr1\Bigg[\left(D_\mu \pp1_{ij}-ig[\Ws{+}_\mu \oms{-}_{ij}-\oms{+}_{ij}\Ws{-}_\mu]+\frac{\pa_\mu \pp0_{ij}}{n_1 e}\right)\nn \\
&&\qquad \times \left(D^\mu \pp1^{ij}-ig[\Ws{+\mu} \oms{-ij}-\oms{+ij}\Ws{-\mu}]+\frac{\pa^\mu \pp0^{ij}}{n_1 e}\right)\Bigg] \nn \\
&&+\frac12\tr1\Bigg[\left(D_\mu \oms{+}_{ij}-ig[\Ws{+}_\mu \pp2_{ij}-\pp1_{ij}\Ws{+}_\mu-e\Ws{+}_\mu \pp0_{ij}]\right)\nn \\
&&\qquad \times\left(D^\mu \oms{-ij}-ig[\Ws{-\mu} \pp1^{ij}-\pp2^{ij}\Ws{+\mu}+e\Ws{+\mu} \pp0_{ij}]\right)\Bigg] \nn \\
&&\frac12\tr2\Bigg[\left(D_\mu \pp2_{ij}-ig[\Ws{-}_\mu \oms{+}_{ij}-\oms{-}_{ij}\Ws{+}_\mu]-\frac{\pa_\mu \pp0_{ij}}{n_2 e}\right)\nn \\
&&\qquad \times \left(D^\mu \pp2^{ij}-ig[\Ws{-\mu} \oms{+ij}-\oms{-ij}\Ws{+\mu}]-\frac{\pa^\mu \pp0^{ij}}{n_2 e}\right)\Bigg] \nn \\
&&+\frac12\tr2\Bigg[\left(D_\mu \oms{-}_{ij}-ig[\Ws{-}_\mu \pp1_{ij}-\pp2_{ij}\Ws{-}_\mu+e\Ws{-}_\mu \pp0_{ij}]\right)\nn \\
&&\qquad \times\left(D^\mu \oms{+ij}-ig[\Ws{+\mu} \pp2^{ij}-\pp1^{ij}\Ws{-\mu}-e\Ws{-\mu} \pp0_{ij}]\right)\Bigg]
\eea
and rearranging
\bea
&&=\frac12\tr1\left(D_\mu \pp1_{ij} D^\mu \pp1^{ij}\right)+\frac12\tr2\left(D_\mu \pp2_{ij} D^\mu \pp2^{ij}\right)
+\frac12 \pa_\mu \pp0_ij \pa^{\mu} \pp0^{ij} \nn \\
&&-ig\tr1\left(D^\mu \pp1^{ij}\left(\Ws{+}_\mu \oms{-}_{ij}-\oms{+}_{ij}\Ws{-}_{ij}\right) \right)-ig\tr2\left(D^\mu \pp2^{ij}\left(\Ws{-}_\mu \oms{+}_{ij}-\oms{-}_{ij}\Ws{+}_{ij}\right) \right) \nn \\
&&-\frac12 ige\left[\tr1(\Ws{+}_\mu \oms{-}_{ij}-\oms{+}_{ij}\Ws{-}_\mu)-\tr2(\Ws{-}_\mu \oms{+}_{ij}-\oms{-}_{ij}\Ws{+}_\mu)\right]\pa^{\mu}\pp0^{ij} \\
&&+\frac12\Bigg[\tr1\left(\left[D_\mu \oms{+}_{ij}-ig(\Ws{+}_\mu \pp2_{ij}-\pp1_{ij} \Ws{+}_\mu)\right]\left[D^\mu \oms{-ij}-ig(\Ws{-\mu}\pp1^{ij} -\pp2^{ij} \Ws{-\mu})\right]\right) \nn \\
&&+\tr2\left(\left[D_\mu \oms{-}_{ij}-ig(\Ws{-}_\mu \pp1_{ij}-\pp2_{ij} \Ws{-}_\mu)\right]\left[D^\mu \oms{+ij}-ig(\Ws{+\mu}\pp2^{ij} -\pp1^{ij} \Ws{+\mu})\right]\right)\Bigg] \nn \\
&&-ige\tr1\left([D_\mu \oms{+}_{ij}-ig(\Ws{+}_\mu \pp2_{ij}-\pp1_{ij}\Ws{+}_\mu)]\Ws{-\mu}\right)\pp0^{ij} \nn \\
&&+ige\tr2\left([D_\mu \oms{-}_{ij}-ig(\Ws{-}_\mu \pp1_{ij}-\pp2_{ij}\Ws{-}_\mu)]\Ws{+\mu}\right)\pp0^{ij} \nn \\
&&+\frac12 g^2e^2\pp0_{ij}\pp0^{ij}\left[\tr1(\Ws{+}_\mu \Ws{-\mu})+\tr2(\Ws{-}_\mu \Ws{+\mu})\right] \nn
\eea
and again, when 2 terms are grouped in a square bracket, they can be condensed into one term by the cyclicity of the trace.  Giving a vev to any component of $\pp0_{ij}$ will clearly induce a mass for the $W$ bosons, as one can see from the last line above.  However, in the second (and third) to last lines, you can see that a 2 point function for $\Ws{+}/\oms{-}$ and $\Ws{-}/\oms{+}$.  This defined the degree of freedom ``eaten'' by the massive gauge field $W$, as we will see later.

The terms of the potential (not expanded) are
\bea
-g\left(\Tr\left(\bar{\tilde{\hat{\chi}}}^i L \hat{\chi}^j \hat{\Phi}_{ij}\right)-\Tr\left(\bar{\hat{\chi}}_i R \tilde{\hat{\chi}}_j\hat{\Phi}^{ij}\right)\right)
\eea
and
\bea
+\frac14 g^2\Tr([\hat{\Phi}_{ij},\hat{\Phi}_{kl}][\hat{\Phi}^{ij},\hat{\Phi}^{kl}])
\eea
where we have defined
\be
\tilde{\hat{\chi}}^I_i=C\left(\overline{\left(\hat{\chi}^{I i}\right)}\right)^T
\ee
where $C$ is the charge conjugation matrix $C=-i\gamma^2\gamma^0$.  In what follows, we will define the operation $\Th$ to be the transpose operation working only on spin indices (as the $T$ above in component notation does).  This is in contrast to $\dagger$, $\overline{\phantom{a}}$, and $*$ which all work on the matrices $\tau^I$ as well.  This allows us to write
\bea
\tilde{\hat{\chi}}^I_i \tau^I&=& C\left(\overline{\left(\hat{\chi}^{I i}\right)}\right)^{\Th} \tau^I \nn \\
&=& C\left(\overline{\left(\hat{\chi}^{I i}\tau^I\right)}\right)^{\Th},
\eea
and so we define
\bea
\tilde{\hat{\chi}}^I_i \tau^I=\tilde{\hat{\chi}}_i&=&\begin{pmatrix}
C\overline{(\cht1^i)}^{\Th} +\frac{C\overline{(\cht0^i)}^{\Th}}{n_1 e} & C\overline{(\xis{-i})}^{\Th} \\
C\overline{(\xis{+i})}^{\Th} & C\overline{(\cht2^i)}^{\Th} -\frac{C\overline{(\cht0^i)}}{n_2 e}
\end{pmatrix} \nn \\
&=&\begin{pmatrix}
C\chb1_i^{\Th} +\frac{C\chb0_i^{\Th}}{n_1 e} & C(\xibs{+}_i)^{\Th} \\
C(\xibs{+}_i)^{\Th} & C\chb2_i^{\Th} -\frac{C\chb0_i^{\Th}}{n_2 e}
\end{pmatrix} \nn \\
& \equiv & \begin{pmatrix}
\cht1_i +\frac{\cht0_i}{n_1 e} & \xits{+}_i \\
\xits{-}_i & \cht2_i - \frac{\cht0_i}{n_2 e}
\end{pmatrix}
\eea
where the last line is read of a definition of symbols.  We further define
\be
\begin{pmatrix}
\overline{\left(\cht1_i\right)} +\frac{\overline{\left(\cht0_i\right)}}{n_1 e} & \overline{\left(\xits{+}_i\right)} \\
\overline{\left(\xits{-}_i\right)} & \overline{\left(\cht2_i\right)} - \frac{\overline{\left(\cht0_i\right)}}{n_2 e}
\end{pmatrix}
\equiv
\begin{pmatrix}
\chtb1^i +\frac{\chtb0^i}{n_1 e} & \xitbs{+i} \\
\xitbs{-i} & \chtb2^i - \frac{\chtb0^i}{n_2 e}
\end{pmatrix}.
\ee
With all these definitions, a few words of clarification is in order.  The way to read the above symbols is simple: the tildes indicate that a charge conjugation has been employed, and the bars that a dirac conjugation.  We have always pulled the indices to the outside of such operations, so that the index structure surrounding the symbol accurately describe its charges under $SU(n_1)\times SU(n_2) \times U(1)$ as well as under the global $SU(4)$.

Using the above definitions, we find
\bea
&&-g\Tr\left(\bar{\tilde{\hat{\chi}}}^i L \hat{\chi}^j \hat{\Phi}_{ij}\right)\nn \\
&&=-g\tr1\Bigg[\left(\chtb1^iL\ch1^j+\xitbs{+i}L\xis{-j}\right)\left(\pp1_{ij}+\frac{\pp0_{ij}}{n_1e}\right)\Bigg] \nn \\
&&-g\tr1\Bigg[\left(\chtb1^iL\xis{+j}+e\chtb0^iL\xis{+j}+\xitbs{+i} L \ch2^j\right)\oms{-}_{ij}\Bigg] \nn \\
&&-g\tr2\Bigg[\left(\chtb2^iL\ch2^j+\xitbs{-i}L\xis{+j}\right)\left(\pp2_{ij}-\frac{\pp0_{ij}}{n_2e}\right)\Bigg] \nn \\
&&-g\tr2\Bigg[\left(\chtb2^iL\xis{-j}-e\chtb0^iL\xis{-j}+\xitbs{-i} L \ch1^j\right)\oms{+}_{ij}\Bigg].
\eea

The other term in the potential can be obtained by replacing spinors without tildes by those with (and vice versa), and switching the $SU(4)$ indices from top to bottom, and finally replacing $L\rightarrow R$.
Therefore, we find
\bea
&&g\Tr\left(\bar{{\hat{\chi}}}_i R \tilde{\hat{\chi}}_j \hat{\Phi}^{ij}\right)\nn \\
&&=g\tr1\Bigg[\left(\chb1_iR\cht1_j+\xibs{+}_iR\xits{-}_j\right)\left(\pp1^{ij}+\frac{\pp0^{ij}}{n_1e}\right)\Bigg] \nn \\
&&+g\tr1\Bigg[\left(\chb1_iR\xits{+}_j+e\chb0_iR\xits{+}_j+\xibs{+}_i R \cht2_j\right)\oms{-ij}\Bigg] \nn \\
&&+g\tr2\Bigg[\left(\chb2_iR\cht2_j+\xibs{-}_iR\xits{+}_j\right)\left(\pp2^{ij}-\frac{\pp0^{ij}}{n_2e}\right)\Bigg] \nn \\
&&+g\tr2\Bigg[\left(\chb2_iR\xits{-}_j-e\chb0_iR\xits{-}_j+\xibs{-}_i R \cht1_j\right)\oms{+ij}\Bigg] \nn \\
\eea

The final term is
\bea
&&\frac14 g^2\Tr([\hat{\Phi}_{ij},\hat{\Phi}_{kl}][\hat{\Phi}^{ij},\hat{\Phi}^{kl}])= \nn \\
&&\frac14 g^2\tr1\Bigg[\left([\pp1_{ij},\pp1_{kl}]+\left[\oms{+}_{ij}\oms{-}_{kl}-\oms{+}_{kl}\oms{-}_{ij}\right]\right)
\left([\pp1^{ij},\pp1^{kl}]+\left[\oms{+ij}\oms{-kl}-\oms{+kl}\oms{-ij}\right]\right)\Bigg] \nn \\
&&\frac14g^2 \tr1\Bigg[\left([\pp1_{ij}\oms{+}_{kl}-\pp1_{kl}\oms{+}_{ij}]+[\oms{+}_{ij}\pp2_{kl}-\oms{+}_{kl}\pp2_{ij}]
+e[\pp0_{ij}\oms{+}_{kl}-\pp0_{kl}\oms{+}_{ij}]\right) \nn \\
&&\qquad \quad \times\left([\pp2^{ij}\oms{-kl}-\pp2^{kl}\oms{-ij}]+[\oms{-ij}\pp1^{kl}-\oms{-kl}\pp1^{ij}]
-e[\pp0^{ij}\oms{-kl}-\pp0^{kl}\oms{-ij}]\right)\Bigg] \nn \\
&&\frac14 g^2\tr2\Bigg[\left([\pp2_{ij},\pp2_{kl}]+\left[\oms{-}_{ij}\oms{+}_{kl}-\oms{-}_{kl}\oms{+}_{ij}\right]\right)
\left([\pp2^{ij},\pp2^{kl}]+\left[\oms{-ij}\oms{+kl}-\oms{-kl}\oms{+ij}\right]\right)\Bigg] \nn \\
&&\frac14 g^2\tr2\Bigg[\left([\pp2_{ij}\oms{-}_{kl}-\pp2_{kl}\oms{-}_{ij}]+[\oms{-}_{ij}\pp1_{kl}-\oms{-}_{kl}\pp1_{ij}]
-e[\pp0_{ij}\oms{-}_{kl}-\pp0_{kl}\oms{-}_{ij}]\right)  \\
&&\qquad \quad \times\left([\pp1^{ij}\oms{+kl}-\pp1^{kl}\oms{+ij}]+[\oms{+ij}\pp2^{kl}-\oms{+kl}\pp2^{ij}]
+e[\pp0^{ij}\oms{+kl}-\pp0^{kl}\oms{+ij}]\right)\Bigg] \nn
\eea

We are now in a position to give a vev to the scalar field $\pp0_{ij}$ to do so, we would like to expand around a vev
\be
\pp0_{ij}=ia_{ij}+\delta\pp0_{ij}
\ee
where $a_{ij}$ are a set of constants with
\be
a_{34}=a_0,\quad a_{12}=-a_0,\quad a^{12}=a_0,\quad a^{34}=-a_0
\ee
with constant $a$ (in what follows we will drop the delta and simply refer to $\pp0_{ij}$).  In the above, the minus signs have been chosen to agree with conditions (\ref{sdeq}).  Now, we expect the Higgs mechanism to transmute certain scalar degrees of freedom into the longitudinal components of the massive $W$ bosons.  This is most easily seen by gauge fixing the scalar sector appropriately. We can read the appropriate gauge fixing by looking for the 2 point function between $\oms{+}$ and $\Ws{-}$. We note that there is a 2 point function between $\pa_\mu(\oms{+}_{12}-\oms{+}_{34})$ and $\Ws{-\mu}$.  Therefore, we define the field
\be
\hat{\Phi}_{12}-\hat{\Phi}_{34}=\hat{\Phi}_{12}
-\left(\hat{\Phi}_{12}\right)^{\dagger}\equiv 2i \hat{P}
\ee
and note that this is an irreducible representation of the gauge group (it is the imaginary component of the complex field).  We now explicitly show that with the above vev, we can gauge away the components of $P$ which are off block diagonal.  We do this by gauge transforming the vev, and showing that off diagonal fluctuations can be generated (and therefore canceled) by such a gauge transformation.  We will show this explicitly for one component.  First, we introduce the notation that
\be
1_{pq}
\ee
is an $n_1$ by $n_2$ matrix with a single non zero entry: a 1 in the $p,q$ position (i.e. $p \in \{1\cdots n_1\},q \in \{1\cdots n_2\}$).  Similarly we define $1_{qp}$.  Therefore, consider the generators
\bea
\delta^1=\frac{1}{\sqrt{2}}\begin{pmatrix} 0 & 1_{pq} \\ 1_{q p} & 0 \end{pmatrix} \nn \\
\delta^2=\frac{i}{\sqrt{2}}\begin{pmatrix} 0 & 1_{pq} \\ -1_{q p} & 0 \end{pmatrix}
\eea
Consider gauge transforming the vev of $P$
\be
<\hat{P}>=-a_0 \tau^0
\ee
via the gauge transformation
\be
-a_0 \tau^0\rightarrow \exp(i\delta^i\lambda^i) (-a_0 \tau^0)\exp(-i\delta^i\lambda^i).
\ee
For infinitesimal $\lambda^i$,
\be
-a_0 \tau^0\rightarrow (1-a_0 i[\delta^i\lambda^i, \tau^0]).
\ee
This commutator is easy to work out, namely
\bea
i[\delta^1,\tau^0]&=&-e\delta^2 \nn \\
i[\delta^2,\tau^0]&=&e\delta^1
\eea
so that finally
\be
-a_0 \tau^0\rightarrow (1-a_0 e(\lambda^2 \delta ^1-\lambda^1 \delta^2).
\ee
Since $\lambda^1$ and $\lambda^2$ are arbitrary and real, we may gauge away completely the real coefficients of the generators $\delta^1$ and $\delta^2$ in $\hat{P}$.  $\hat{P}$ is a Hermetian field, and so the coefficents of $\delta^1$ and $\delta^2$ are real, and so finally their coefficients (in $\hat{P}$) may be completely removed with an appropriate gauge choice.  This statement is obviously true for all off diagonal components of $\hat{P}$ and so we find that $\oms{+}_{12}-\oms{+}_{34}=0$ is an appropriate gauge choice.  Statements of this gauge fixing may be written in the following useful forms
\bea
a^{ij}\oms{\pm}_{ij}=0 \\
\oms{\pm}_{12}=\oms{\pm}_{34}=\left(\oms{\mp}_{12}\right)^{\dagger}
\eea
some of which will directly appear in the action when we expand.  Above we have noted that one can compare to the $\oms{-}$ statements by taking a $\dagger$ of the original equations, and replacing ${12}\leftrightarrow {34}$.
The particular case of $SU(2)$ is studied in \cite{Schabinger:2008ah}.  Discussion of the appropriate massive representations of the $\mathcal N=4$ supersymmetry group was studied in \cite{Fayet:1978ig}.

\section{On different IR regulators}\label{appendix:regularization}
Here we will consider the two different regulations of the Kruczenski ``wedge'' solution and show that their divergences cancel, and so the leading order IR divergences indeed cancel.  The two regulators considered are
\begin{enumerate}
\item A strict radial cutoff, taking $r\in [\epsilon, \Lambda]$.  We will want to consider the divergences as $\epsilon \rightarrow 0$ and $\Lambda \rightarrow \infty$ with the solution obeying boundary conditions set at $r=0$.
\item A modification where the boundary conditions are set at $r=\epsilon$ (and the integration ends here as well).  This cures IR divergences, but not the UV ones, and so we will still need a UV cutoff $\Lambda$.
\end{enumerate}

We begin with a word of warning.  One must always be careful when regulating integrals using coordinate transformations that are functions of the limits of integration.  Such coordinate transformations can ``shuffle infinities'' to make certain IR divergences appear to be UV ones, and vice versa.  We will use the wedge solution to illustrate this point in a concrete manner.

To do so, we will first display the wedge solution, and its counterpart with boundary conditions set at $r=\epsilon$.  The boundary conditions are boost invariant, and in the case where the cusp is at the boundary $r=0$, it is also scale invariant.  However, since the second set of boundary conditions we wish to consider break the scale symmetry, we enforce the boost invariance on the solution only.  Therefore, we look for solutions of the form
\be
x^{\pm}=\exp(\tau \pm \sigma), \quad r=\exp(\tau)w(\tau).
\ee
The action then reduces to \cite{Alday:2007hr,Kruczenski:2002fb}
\be
\int d\tau \frac{\sqrt{(w'+w)^2-1}}{w^2}.
\ee
Given that the action is explicitly $\tau$ independent, we may write the associated first integral \cite{Berkovits:2008ic}
\be
c=\frac{w(w'+w)-1}{w^2\sqrt{(w'+w)^2-1}},
\ee
which one can solve as
\be
w'=-\frac{-(1-w^2+c^2w^4)+cw\sqrt{1-w^2+c^2w^4}}{w(c^2w^2-1)}.
\ee
Looking for constant solutions, one finds that
\be
w=\sqrt{2}
\ee
is a solution with
\be
c=\frac12.
\ee
This is the solution with the ``wedge'' boundary conditions satisfied at $r=0$.  The solution with boundary conditions at $r=\epsilon$ was found in \cite{Berkovits:2008ic} and is given implicitly by
\be
e^{\tau}=\epsilon\left(\frac{w+\sqrt{2}}
{w-\sqrt{2}}\right)^{\frac{1}{\sqrt{2}}}\frac{1}{1+w}.
\ee
The above solution asymptotes to the original wedge when $\tau\rightarrow \infty$ which is equivalently $w\rightarrow \sqrt{2}$.  The above has the first integral $c=\frac12$ as with the original wedge.

We now illustrate the warning above.  Let us consider the usual wedge solution given by $w=\sqrt{2}$.  In this case, the regulated action is
\be
\int_{\ln\left(\frac{\epsilon}{\sqrt{2}}\right)}
^{\ln\left(\frac{\Lambda}{\sqrt{2}}\right)} d\tau=\ln\left(\frac{\Lambda}{\sqrt{2}}\right)
-\ln\left(\frac{\epsilon}{\sqrt{2}}\right)
\ee
Now consider the coordinate transformation
\be
e^{\tau}=\epsilon \frac{w}{\sqrt{2}(w+1)}
\left(\frac{w+\sqrt{2}}{w-\sqrt{2}}\right)^{\frac{1}{\sqrt{2}}}.
\ee
This transforms our integral to the form
\be
\int_{w_{\Lambda,\epsilon}}^{\infty} dw \frac{w^2+2w+2}{2w(w+1)(w^2-2)}
\ee
where $w_{\Lambda,\epsilon}$ is the solution to
\be
\frac{\Lambda}{\sqrt{2}}=\epsilon \frac{w_{\Lambda,\epsilon}}{\sqrt{2}(w_{\Lambda,\epsilon}+1)}
\left(\frac{w_{\Lambda,\epsilon}+\sqrt{2}}
{w_{\Lambda,\epsilon}-\sqrt{2}}\right)^{\frac{1}{\sqrt{2}}}.
\ee
Note that the above integrand converges as $w^{-2}$ for $w\rightarrow \infty$.  This, however, does not mean that there is no IR divergence.  This is because $\epsilon$ explicitly appears in the other bound of integration $w_{\Lambda,\epsilon}$.  Recall that for large $\Lambda$, $w_{\Lambda,\epsilon}$ is close to $\sqrt{2}$, and $z$ is large.  One would have associated this with a UV divergence, but because we have made a coordinate transformation that explicitly uses the IR regulator, the finial regulator $w_{\Lambda,\epsilon}$ depends explicitly on this.  Further, $w_{\Lambda,\epsilon}$ is a function only of $\Lambda/\epsilon$, which we can see both from its defining equation, and also from the answer in the more trivial coordinates (which it has to match).

We wish to play a similar game for the solution where the cusp has been moved to the location $r=\epsilon$.  Using the integral of motion $c$, and the resulting expression for $w'$ we may write the action in terms of $w$ and $d\tau$.  Further, the implicit relation between $w$ and $\tau$ gives $d\tau$ in terms of $w$ and $dw$, so we may write the action in this case as
\be
\int_{w_{\Lambda,\epsilon}}^\infty dw
\frac{2\sqrt{w^6+3 w^5-6w^3-3w^2+2w+1}}{w^2(w+1)(w^2-2)}.
\ee
Now we can see why the previous coordinate transformation was important.  The new solution for $r$ written in $w$ coordinates is
\be
r=\epsilon \frac{w}{(w+1)}
\left(\frac{w+\sqrt{2}}{w-\sqrt{2}}\right)^{\frac{1}{\sqrt{2}}}.
\ee
and so the $w_{\Lambda,\epsilon}$ is the same as that defined earlier, given that we are making strict $r$ cutoffs.  The difference in actions can now be written
\be
\int_{w_{\Lambda,\epsilon}}^{\infty}dw\left( \frac{2\sqrt{w^6+3 w^5-6w^3-3w^2+2w+1}}{w^2(w+1)(w^2-2)}-\frac{w^2+2w+2}{2w(w+1)(w^2-2)}\right).
\ee
There is no pole at $w=\sqrt{2}$ in the above integral: it cancels.  This means that all divergences between the two actions have canceled.  In particular, that means that they are divergent as the same function of $\Lambda/\epsilon$ with $\Lambda$ and $\epsilon$ defined the same way.  Hence, they are actually divergent in the same way in either $\epsilon$ or $\Lambda$.  Thus, we have canceled both the UV and IR divergences.\footnote{  In this discussion, we have neglected the divergence associated with the integral over $\sigma$.}  This, however, does not mean that we have an unambiguous answer.  Recall that when subtracting infinities, one may arrive at any constant.  Hence, the above constant (after integration) is by no means special.

Below we outline the calculation.  To lend clarity below, we call $\omega(\tau)$ the function of tau, reserving $w$ for the change of variables $\tau(w)$.

\begin{center}
\begin{tabular}{c|c|c}
& Solution 1: Cusp at $r=0$ & Solution 2: Cusp at $r=\epsilon$ \\
\hline
form of sln. & $x^{\pm}=\exp(\tau \pm \sigma), r=\exp(\tau)\omega(\tau)$ & $x^{\pm}=\exp(\tau \pm \sigma), r=\exp(\tau)\omega(\tau)$ \\
\hline
form of act. & $\int d\tau \frac{\sqrt{(\omega'+\omega)^2-1}^{\phantom{I}}}{\omega^2}$ & $\int d\tau \frac{\sqrt{(\omega'+\omega)^2-1}^{\phantom{I}}}{\omega^2}$ \\
\hline
first integral & $c=\frac{\omega(\omega'+\omega)-1}{\omega^2\sqrt{(\omega'+\omega)^2-1}}$ & $c=\frac{\omega(\omega'+\omega)-1}{\omega^2\sqrt{(\omega'+\omega)^2-1}}$\\
\hline
solve for $\omega'$ & $\omega'=-\frac{-(1-\omega^2+c^2\omega^4)
+c\omega\sqrt{1-\omega^2+c^2\omega^4}^{\phantom{I}}}{\omega(c^2\omega^2-1)}$ & $\omega'=-\frac{-(1-\omega^2+c^2\omega^4)
+c\omega\sqrt{1-\omega^2+c^2\omega^4}^{\phantom{I}}}{\omega(c^2\omega^2-1)}$ \\
\hline
Solution & $\omega=\sqrt{2}$ & $e^{\tau}=\epsilon\left(\frac{\omega+\sqrt{2}}
{\omega-\sqrt{2}}\right)^{\frac{1}{\sqrt{2}}}\frac{1}{1+\omega}$ \\
\hline
value of $c$& $\frac12$ & $\frac12$ \\
\hline
$\tau(w)$ coord. & $e^{\tau}=\epsilon\left(\frac{w+\sqrt{2}}
{w-\sqrt{2}}\right)^{\frac{1}{\sqrt{2}}}\frac{w}{\sqrt{2}(1+w)}$ &$e^{\tau}=\epsilon\left(\frac{w+\sqrt{2}}
{w-\sqrt{2}}\right)^{\frac{1}{\sqrt{2}}}\frac{1}{1+w}$ \\
\hline
action & $\int_{w_{\Lambda,\epsilon}}^{\infty}dw \frac{w^2+2w+2}{2w(w+1)(w^2-2)}$ &  $\int_{w_{\Lambda,\epsilon}}^{\infty}dw \frac{2\sqrt{w^6+3 w^5-6w^3-3w^2+2w+1}^{\phantom{I}}}{w^2(w+1)(w^2-2)}$
\end{tabular}
\end{center}
with $w_{\Lambda,\epsilon}$ defined by
\be
\frac{\Lambda}{\sqrt{2}}=\epsilon \frac{w_{\Lambda,\epsilon}}{\sqrt{2}(w_{\Lambda,\epsilon}+1)}
\left(\frac{w_{\Lambda,\epsilon}+\sqrt{2}}
{w_{\Lambda,\epsilon}-\sqrt{2}}\right)^{\frac{1}{\sqrt{2}}}
\ee
as before.

\end{document}